\documentstyle[aps,preprint,eqsecnum]{revtex}
\begin{document}
\draft
 
{\tighten

\title{Quantum Monte Carlo calculations of nuclei with $A\leq7$}
\author{B. S. Pudliner\cite{bsp} and V. R. Pandharipande\cite{vrp}}
\address{Physics Department, University of Illinois at Urbana-Champaign,
         1110 West Green St., Urbana, Illinois 61801}
\author{J. Carlson\cite{jc}}
\address{Theoretical Division, Los Alamos National Laboratory, Los Alamos
         New Mexico 87545}
\author{Steven C. Pieper\cite{scp} and R. B. Wiringa\cite{rbw}}
\address{Physics Division, Argonne National Laboratory, Argonne, IL 60439}
\date{\today}
\maketitle
 
\begin{abstract}
We report quantum Monte Carlo calculations of ground and low-lying excited 
states for nuclei with $A\leq7$ using a realistic Hamiltonian containing the 
Argonne $v_{18}$ two-nucleon and Urbana IX three-nucleon potentials.
A detailed description of the Green's function Monte Carlo algorithm
for systems with state-dependent potentials is given
and a number of tests of its convergence and accuracy are performed.
We find that the Hamiltonian being used results in ground states of 
both $^6$Li and $^7$Li that are stable against 
breakup into subclusters, but somewhat underbound compared to experiment.
We also have results for $^6$He, $^7$He, and their isobaric analogs.
The known excitation spectra of all these nuclei are reproduced reasonably well
and we predict a number of excited states in $^6$He and $^7$He.
We also present spin-polarized one-body and several different two-body 
density distributions.
These are the first microscopic calculations that directly produce nuclear shell
structure from realistic interactions that fit $N\!N$ scattering data.
\end{abstract}
 
\pacs{PACS numbers: 21.10.-k, 21.45+v, 21.60.Ka}

}
 
\narrowtext

\section{INTRODUCTION}

A major goal in nuclear physics is to understand how nuclear binding, 
stability, and structure arise from the underlying interactions 
between individual nucleons.
A large amount of empirical information about the nucleon-nucleon 
scattering problem has been accumulated over time, resulting in ever more 
sophisticated $N\!N$ potential models.
However, for many years, it has been feasible to calculate exactly only 
three- and four-body nuclei with realistic $N\!N$ interactions.
Thanks to the ongoing advances in computational resources, 
particularly the advent of massively parallel computers, it is now possible
to apply sophisticated quantum Monte Carlo methods to the study of light
$p$-shell nuclei, which have a far richer spectrum to investigate.

The $p$-shell has long been a testing ground for shell model methods~\cite{CK65}.
Shell-model studies of $A=4-7$ nuclei have recently progressed to the
stage of large-basis, ``no-core'' calculations using $G$-matrices derived
from the latest $N\!N$-interaction models~\cite{ZBVHS95}.  
Alternatively, there have been extensive studies of these light nuclei
with cluster-cluster models, using combinations of $N\!N$ and $\alpha\!N$
potentials~\cite{KPRVR95}.
Our goal here is to calculate properties, in particular ground-state energies,
of the light $p$-shell nuclei directly from bare $N\!N$ and $N\!N\!N$ interactions,
without any intermediate effective interaction.

Recently we reported~\cite{PPCW95} first results for the ground states of 
$^6$He, $^6$Li, and $^6$Be, and the first two excited states in $^6$Li, 
calculated with the Green's function Monte Carlo (GFMC) method for a 
Hamiltonian containing the Argonne $v_{18}$ $N\!N$ and Urbana IX $N\!N\!N$ 
potentials.
Here we present improved and extended calculations for the $A=6$ nuclei
and the first detailed studies of $^7$He, $^7$Li, $^7$Be, and $^7$B,
including all the observed low-lying excitations.
In this work, it is possible to see for the first time the full splendor
of the nuclear shell structure emanating directly from a microscopic 
interaction that fits $N\!N$ scattering data.

The Argonne $v_{18}$ model~\cite{WSS95} is one of a class of new, highly
accurate $N\!N$ potentials that fit both $pp$ and $np$ scattering data up to 
350 MeV with a $\chi^2/$datum near 1.
This necessarily involves the introduction of charge-independence breaking
in the strong force; a complete electromagnetic interaction is also
included as an integral part of the model.
This makes the model useful for studying charge dependence and charge-symmetry
breaking in nuclei.
The $N\!N$ potential is supplemented by a three-nucleon interaction from the
Urbana series of $N\!N\!N$ potentials~\cite{CPW83}, including both long-range
two-pion exchange and a short-range phenomenological component.
The Urbana model IX is adjusted to reproduce the binding energy of $^3$H
and give a reasonable saturation density in nuclear matter when used with 
Argonne $v_{18}$.
Details of the Hamiltonian are given in Sec.II.

The first step in our calculation is the construction of suitable trial
functions.
Variational wave functions based on products of correlated operators have been
used successfully for $^3$H and $^4$He, giving binding energies about 2\% above 
exact Faddeev, hyperspherical harmonic, or GFMC solutions~\cite{APW95}.
We generalize this type of trial function for $A=6,7$ nuclei, adding $p$-wave 
orbitals and using $LS$ coupling to produce all possible $(J^{\pi};T)$ 
quantum states.
Parameters in the trial functions are adjusted to minimize the energy
expectation value, evaluated with Monte Carlo integration, subject to the 
constraint that rms radii are close to the experimental values for $^6$Li 
and $^7$Li.
Unfortunately, the best trial functions we have been able to build do not
give $p$-shell nuclei stable against breakup into subclusters.
Nevertheless, these trial functions provide a good starting point for the
GFMC calculation. 
The variational wave functions and a brief description of the variational
Monte Carlo (VMC) calculations are given in Sec.III.

The GFMC method projects out the exact lowest energy state, $\Psi_0$, for a 
given set of quantum numbers, from a suitable trial function, $\Psi_T$, using 
$\Psi_0 = \lim_{\tau \rightarrow \infty} \exp [ - ( H - E_0) \tau ] \Psi_T$.
The method has been used with great success in a variety of condensed matter 
problems, and in $s$-shell nuclei with realistic interactions~\cite{C88}.
Our first calculations for $p$-shell nuclei~\cite{PPCW95} were made with a 
short time approximation for the propagation in imaginary time, 
carried out to $\tau = 0.06$ MeV$^{-1}$ and extrapolated to $\tau = \infty$.
In the present work we have improved the algorithm by adopting an exact 
two-body propagator which allows bigger time steps, saving significantly
on the computational cost.
We also have started our calculations with better trial functions, which 
allows more reliable results to be obtained from GFMC propagations that are
limited to small $\tau$.
A detailed discussion of the method as applied to realistic nuclear
forces is given in Sec.IV.

The VMC and GFMC calculations for $p$-shell nuclei are very computer intensive, 
and would not have been possible without the recent advances in computational 
power due to the advent of massively parallel machines.
Section V describes the implementation of the GFMC algorithm in a parallel
environment, including issues of communication between processors and
load-balancing.
We have also performed a number of tests of the GFMC method, including
comparisons to other exact calculations for $s$-shell nuclei, studies of 
extrapolation in $\tau$, and sensitivity to the quality of the input trial
function.
These tests are described in Sec.VI.

Results of our GFMC calculations are presented in Sec.VII.
We have obtained energies for five states of unique $(J^{\pi};T)$ in the
$A=6$ nuclei, and another five states in $A=7$ nuclei, not counting isobaric
analog states.
In general we find the nuclei are slightly underbound with the present
Hamiltonian, but $^6$Li ($^7$Li) is stable against breakup into 
$\alpha + d$ ($\alpha + t$).
The low-lying excited states are correctly ordered with reasonable excitation 
energies.
The VMC energies are found to lie $\sim 3$ (4.5) MeV above the GFMC results
in $A=6$ (7) nuclei, but with very similar excitation energies.
We have also used the VMC wave functions to perform small-basis
diagonalizations of states with the same quantum numbers but
different symmetries.  These calculations optimize the admixtures
of different-symmetry contributions to the ground state, and also
provide estimates for higher-lying
excited states with the same $(J^{\pi};T)$ quantum numbers.
We verify that these states, not all of which have been observed, do indeed
lie at moderately higher excitations.
The VMC spectra are also discussed in Sec.VII.

One- and two-body density distributions from both VMC and GFMC calculations
are presented in Sec.VIII.
These include the densities of spin up/down nucleons in polarized $^6$Li and 
$^7$Li.
In general, the GFMC densities are slightly more peaked than the input VMC
densities, but overall they are very similar.
Unfortunately, both VMC and GFMC calculations are not very sensitive to
the very long range properties of the wave functions.  
Therefore it is not yet possible to accurately calculate the quadrupole 
moments and asymptotic $D/S$ ratios with these methods.
Finally, we present our conclusions in Sec.IX.

\section{HAMILTONIAN}

Our Hamiltonian includes a nonrelativistic one-body kinetic energy, the
Argonne $v_{18}$ two-nucleon potential \cite{WSS95} and the Urbana IX
three-nucleon potential \cite{PPCW95},
\begin{equation}
   H = \sum_{i} K_{i} + \sum_{i<j} v_{ij} + \sum_{i<j<k} V_{ijk} \ .
\end{equation}
The kinetic energy operator has charge-independent (CI) and 
charge-symmetry-breaking (CSB) components, the latter due to 
the difference in proton and neutron masses,
\begin{equation}
   K_{i} = K^{CI}_{i} + K^{CSB}_{i} \equiv -\frac{\hbar^2}{4} 
     (\frac{1}{m_{p}} + \frac{1}{m_{n}}) \nabla^{2}_{i}
    -\frac{\hbar^2}{4} (\frac{1}{m_{p}} - \frac{1}{m_{n}})\tau_{zi} 
     \nabla^{2}_{i} ~.
\end{equation}

The Argonne $v_{18}$ potential can be written as a sum of electromagnetic  and 
one-pion-exchange terms and a shorter-range phenomenological part,
\begin{equation}
   v_{ij} = v^{\gamma}_{ij} + v^{\pi}_{ij} + v^{R}_{ij} \ .
\end{equation}
The electromagnetic terms include one- and two-photon-exchange Coulomb 
interaction, vacuum polarization, Darwin-Foldy, and magnetic moment terms,
with appropriate proton and neutron form factors:
\begin{eqnarray}
 &&  v^{\gamma}(pp) = V_{C1}(pp) + V_{C2} + V_{VP} + V_{DF} + V_{MM}(pp) \ , 
\label{eq:vempp} \\
 &&  v^{\gamma}(np) = V_{C1}(np) + V_{MM}(np) \ , \\
 &&  v^{\gamma}(nn) = V_{MM}(nn) \ .
\label{eq:vemnn}
\end{eqnarray}
The $V_{MM}$ contain spin-spin, tensor, and spin-orbit components.
Detailed expressions for these terms, including the form factors, are
given in Ref.~\cite{WSS95}.  
These terms should be included in calculations aiming for better than 99\% 
accuracy.  
For example, the contribution of $v^{\gamma}$(np) to the binding energy of 
the deuteron is $\sim$ 0.02 MeV, i.e. 1\% of the total.

The one-pion-exchange part of the potential includes the charge-dependent (CD) 
terms due to the difference in neutral and charged pion masses.
It can be written in an operator format as
\begin{equation}
   v^{\pi}_{ij} = f^{2} \left(\frac{m}{m_{s}}\right)^{2} \case{1}{3}
                 mc^{2} \left[ X_{ij} \tau_{i} \cdot \tau_{j} 
                      + \tilde{X}_{ij} T_{ij} \right] \ ,
\end{equation}
where $T_{ij} = 3\tau_{zi}\tau_{zj}-\tau_{i}\cdot\tau_{j}$ is the isotensor
operator and
\begin{eqnarray}
   && X_{ij} = \case{1}{3} \left( X^{0}_{ij} + 2 X^{\pm}_{ij} \right) , \\
   && \tilde{X}_{ij} = \case{1}{3} \left( X^{0}_{ij} - X^{\pm}_{ij} 
                                      \right) , \\
   && X^{m}_{ij} =  \left[ Y(mr) \sigma_{i} \cdot \sigma_{j} +
                           T(mr) S_{ij}  \right] \ .
\end{eqnarray}
Here $Y(mr)$ and $T(mr)$ are the normal Yukawa and tensor functions
with a cutoff specified in Ref.~\cite{WSS95}, and $X^{\pm,0}$ are calculated
with $m = m_{\pi^{\pm}}$ and $m_{\pi^{0}}$.

The one-pion-exchange and the remaining phenomenological part of the potential 
can be written as a sum of eighteen operators,
\begin{equation}
       v^{\pi}_{ij} + v^{R}_{ij} = \sum_{p=1,18} v_{p}(r_{ij}) O^{p}_{ij} \ .
\end{equation}
The first fourteen are charge-independent, 
\begin{eqnarray}
O^{p=1,14}_{ij} = [1, ({\bf\sigma}_{i}\cdot{\bf\sigma}_{j}), S_{ij},
({\bf L\cdot S}),{\bf L}^{2},{\bf L}^{2}({\bf\sigma}_{i}\cdot{\bf\sigma}_{j}),
({\bf L\cdot S})^{2}]\otimes[1,({\bf\tau}_{i}\cdot{\bf\tau}_{j})] ~,
\end{eqnarray}
and the last four,
\begin{equation}
   O^{p=15,18}_{ij} = [1, ({\bf\sigma}_{i}\cdot{\bf\sigma}_{j}), 
S_{ij}]\otimes T_{ij} , (\tau_{zi}+\tau_{zj}) \ ,
\end{equation}
break charge independence.
We will refer to the potential from the $p=15-17$ terms as $v^{CD}$ and from
the $p=18$ term as $v^{CSB}$.
We note that in the context of isospin symmetry the CI, CSB
and CD terms are respectively isoscalar, isovector and isotensor.

The potential was fit directly to the Nijmegen $N\!N$ scattering data
base~\cite{BCKKRSS90,SKRS93} containing 1787 $pp$ and 2514 $np$ data in
the range $0-350$ MeV, with a $\chi^2$ per datum of 1.09.
It was also fit to the $nn$ scattering length measured in $d(\pi^-,\gamma)nn$ 
experiments and the deuteron binding energy.
There could in principle be more charge-independence-breaking (CIB) terms 
such as ${\bf L}\cdot{\bf S} T_{ij}$ or $S_{ij} (\tau_{zi}+\tau_{zj})$ but 
the scattering data are not sufficiently precise to identify them at present.

The Urbana series of three-nucleon potentials is written as a sum of 
two-pion-exchange and shorter-range phenomenological terms,
\begin{equation}
   V_{ijk} = V^{2\pi}_{ijk} + V^{R}_{ijk} ~.
\end{equation}
The two-pion-exchange term can be expressed simply as
\begin{equation}
   V^{2\pi}_{ijk} = \sum_{cyclic} A_{2\pi} 
           \{X^{\pi}_{ij},X^{\pi}_{jk}\} 
           \{\tau_{i}\cdot\tau_{j},\tau_{j}\cdot\tau_{k}\}
         + C_{2\pi} [X^{\pi}_{ij},X^{\pi}_{jk}]
           [\tau_{i}\cdot\tau_{j},\tau_{j}\cdot\tau_{k}] \ ,
\end{equation}
where $X^{\pi}_{ij}$ is constructed with an average pion mass, 
$m_{\pi}=\case{1}{3}m_{\pi^0}+\case{2}{3}m_{\pi^{\pm}}$.
The anticommutator and commutator terms are denoted by $V^{A}_{ijk}$ and
$V^{C}_{ijk}$, respectively, and for the Urbana models
$C_{2\pi} = \case{1}{4}A_{2\pi}$, as in the original 
Fujita-Miyazawa model~\cite{FM57}.
The shorter-range phenomenological term is given by
\begin{equation}
   V^{R}_{ijk} = \sum_{cyclic} U_0 T^2(m_{\pi}r_{ij}) T^2(m_{\pi}r_{jk}) \ .
\label{eq:vrijk}
\end{equation}
The parameters for model IX are $A_{2\pi} = -0.0293$ MeV and $U_0 = 0.0048$ MeV.
They have been determined by fitting the density of nuclear matter and the 
binding energy of $^3$H in conjunction with the Argonne $v_{18}$ interaction.  
These values are only slightly different from the model VIII values, 
$A_{2\pi} = -0.028$ MeV and $U_0 = 0.005$ MeV, that were adjusted for use
with the older Argonne $v_{14}$ interaction.
In principle, the $V^{R}_{ijk}$ can have other terms~\cite{CPR95}, however 
we need additional data to obtain their strengths; presumably a part of it 
is due to relativistic effects~\cite{BWBS87,FPF95,FPCR95}.

Direct GFMC calculations with the full interaction (in particular
spin-dependent terms which involve the square of the momentum operator)
have very large statistical errors,
for reasons that will be discussed in section IV.
Also the CIB terms in $H$ are fairly weak and therefore can be treated 
conveniently as a first order perturbation.
Further, using a wave function of good isospin significantly reduces 
the cost of the calculations. 
Hence we construct the GFMC propagator with a simpler isoscalar 
Hamiltonian,
\begin{equation}
     H^{\prime} = \sum_{i}K^{CI}_{i} + \sum_{i<j} v^{\prime}_{ij} 
                + \sum_{i<j<k}V^{\prime}_{ijk} \ ,
\label{eq:hgfmc}
\end{equation}
where $v^{\prime}_{ij}$ is defined as
\begin{equation}
     v^{\prime}_{ij}=\sum_{p=1,8} v^{\prime}_{p}(r_{ij}) O^{p}_{ij}
                                + v^{\prime}_{C}(r_{ij}) \ .
\label{eq:vprime}
\end{equation}
The interaction $v^{\prime}_{ij}$ has only eight operator terms, with operators
$[1, ({\bf\sigma_{i}}\cdot{\bf\sigma_{j}}), S_{ij},({\bf L\cdot S})]\otimes[1,({\bf\tau_{i}}\cdot{\bf\tau_{j}})]$, chosen such that it equals the isoscalar
part of the full interaction in all $S$ and $P$ waves as well as in the
$^{3}D_{1}$ wave and its coupling to the $^{3}S_{1}$.
The strong interaction terms are related to the full $v_{ij}$ by
\begin{eqnarray}
 v^{\prime}_{1} &=& v_1 + \case{5}{4}v_9 + \case{3}{4}v_{10} + \case{3}{4}v_{11}
  + \case{9}{4}v_{12} + \case{3}{4}v_{13} + \case{3}{4}v_{14} ~, \nonumber \\
 v^{\prime}_{2} &=& v_2 + \case{1}{4}v_9 + \case{3}{4}v_{10} + \case{3}{4}v_{11}
  - \case{3}{4}v_{12} + \case{1}{4}v_{13} + \case{1}{4}v_{14} ~, \nonumber \\
 v^{\prime}_{3} &=& v_3 + \case{1}{4}v_9 + \case{3}{4}v_{10} + \case{3}{4}v_{11}
  - \case{3}{4}v_{12} + \case{1}{4}v_{13} + \case{1}{4}v_{14} ~, \nonumber \\
 v^{\prime}_{4} &=& v_4 + \case{1}{4}v_9 - \case{1}{4}v_{10} - \case{1}{4}v_{11}
  + \case{5}{4}v_{12} + \case{1}{12}v_{13} + \case{1}{12}v_{14} ~, \nonumber \\
 v^{\prime}_{5} &=& v_5 - \case{5}{16}v_{13} - \case{5}{16}v_{14} ~, \nonumber \\
 v^{\prime}_{6} &=& v_6 - \case{5}{48}v_{13} - \case{5}{48}v_{14} ~, \nonumber \\
 v^{\prime}_{7} &=& v_7 - \case{1}{2}v_9 + \case{3}{2}v_{10} - \case{1}{2}v_{11}
  + \case{3}{2}v_{12} - \case{9}{8}v_{13} + \case{15}{8}v_{14} ~, \nonumber \\
 v^{\prime}_{8} &=& v_8 + \case{1}{2}v_9 - \case{3}{2}v_{10} + \case{1}{2}v_{11}
  - \case{3}{2}v_{12} + \case{5}{8}v_{13} - \case{19}{8}v_{14} \ .
\end{eqnarray}

The isoscalar part of $V_{C1}(pp)$ is also included in $H^{\prime}$.
We derive it by writing the projector for a pair of
protons in terms of isoscalar, isovector, and isotensor operators:
\begin{equation}
     \case{1}{4}(1+\tau_{zi})(1+\tau_{zj}) = \case{1}{4}(1 + \case{1}{3} \tau_i\cdot\tau_j
    + \tau_{zi}+\tau_{zj} + \case{1}{3}T_{ij})  \  ,
\label{eq:ppop}
\end{equation}
\begin{equation}
     v^{\prime}_{C}(r_{ij}) = [\alpha_C(A,Z,T) + \case{1}{12}\tau_i\cdot\tau_j]
                              V_{C1}(pp) \ ,
\label{eq:vcprime}
\end{equation}
\begin{equation}
     \alpha_C(A,Z,T) = \case{1}{A(A-1)} [Z(Z-1) + \case{1}{4}A - \case{1}{3}T(T+1)] \ .
\label{eq:alphac}
\end{equation}
The sum over all pairs of $\alpha_C(A,Z,T) + \case{1}{12}\tau_i\cdot\tau_j$
is just the number of $pp$ pairs in the given nucleus.

The $v^{\prime}_{ij}$ is a little more attractive than $v_{ij}$; for example,
$^4$He is overbound by $\sim 2$ MeV with $v^{\prime}_{ij}$.
The expectation value of the difference $\langle v_{ij}-v^{\prime}_{ij} \rangle$
scales like $\langle V_{ijk} \rangle$, presumably because three-body and 
higher-order clusters give important contributions to it.
Note that in $^3$H and $^4$He the two-body cluster gives zero contribution to
$\langle v_{ij}-v^{\prime}_{ij} \rangle$ since they are identical in
low partial waves.
We compensate for this tendency towards overbinding by using a 
$V^{\prime}_{ijk}$ in which the repulsive $U_0$ term of Eq.(\ref{eq:vrijk})
has been increased by $\sim 30\%$ in the $H^{\prime}$.
This ensures $\langle H^{\prime} \rangle \approx \langle H \rangle$ so that
the GFMC propagation does not produce excessively large densities due to 
overbinding.
The contribution of $(H-H^{\prime})$ is calculated perturbatively.

\section{VARIATIONAL MONTE CARLO}

The variational method can be used to obtain approximate solutions to the 
many-body Schr\"{o}dinger equation, $H\Psi = E\Psi$, for a wide range of 
nuclear systems, including few-body nuclei, light closed shell nuclei, 
nuclear matter, and neutron stars~\cite{W93}.
A suitably parameterized wave function, $\Psi_V$, is used to calculate an
upper bound to the exact ground-state energy,
\begin{equation}
   E_V = \frac{\langle \Psi_V | H | \Psi_V \rangle}
              {\langle \Psi_V   |   \Psi_V \rangle} \geq E_0 \ .
\label{eq:expect}
\end{equation}
The parameters in $\Psi_V$ are varied to minimize $E_V$, and the lowest value
is taken as the approximate ground-state energy. 

Upper bounds to excited states are also obtainable, either
from standard VMC calculations if they have different quantum
numbers from the ground state, or from small-basis diagonalizations
if they have the same quantum numbers.
The corresponding $\Psi_V$ can then be used to calculate other properties,
such as particle density or electromagnetic moments, or it can be used as 
the starting point for a Green's function Monte Carlo calculation.
In this section we first describe our {\it ansatz} for $\Psi_V$ for the light
$p$-shell nuclei and then briefly review how the expectation value, 
Eq.(\ref{eq:expect}), is evaluated and the parameters of $\Psi_V$ are fixed.

\subsection{Wave Function}

Our best variational wave function for the nuclei studied here has the 
form~\cite{APW95}
\begin{equation}
     |\Psi_V\rangle = \left[1 + \sum_{i<j<k}(U_{ijk}+U^{TNI}_{ijk}) 
                              + \sum_{i<j}U^{LS}_{ij} \right]
                      |\Psi_P\rangle \ ,
\label{eq:bestpsiv}
\end{equation}
where the pair wave function, $\Psi_P$, is given by
\begin{equation}
     |\Psi_P\rangle = \left[ {\cal S}\prod_{i<j}(1+U_{ij}) \right] 
                      |\Psi_J\rangle \ .
\label{eq:psip}
\end{equation}
The $U_{ij}$, $U^{LS}_{ij}$, $U_{ijk}$, and $U^{TNI}_{ijk}$ are noncommuting 
two- and three-nucleon correlation operators, and the ${\cal S}$ is a
symmetrization operator.
The form of the totally antisymmetric Jastrow wave function, 
$\Psi_J$, depends on the nuclear state under investigation.
For the $s$-shell nuclei we use the simple form
\begin{equation}
     |\Psi_J\rangle = \left[ \prod_{i<j<k}f^c_{ijk} \right]
                      \left[ \prod_{i<j}f_c(r_{ij}) \right] 
                     |\Phi_A(JMTT_{3})\rangle \ .
\label{eq:jastrow}
\end{equation}
Here $f_c(r_{ij})$ and $f^c_{ijk}$ are central two- and three-body correlation
functions and
\begin{eqnarray}
 &&  |\Phi_{3}(\case{1}{2} \case{1}{2} \case{1}{2} \case{1}{2}) \rangle
        = {\cal A} |p\uparrow p\downarrow n\uparrow \rangle \ , \\
 &&  |\Phi_{4}(0 0 0 0) \rangle
        = {\cal A} |p\uparrow p\downarrow n\uparrow n\downarrow \rangle \ .
\end{eqnarray}

The two-body correlation operator $U_{ij}$ is a sum of spin, isospin, and
tensor terms:
\begin{equation}
     U_{ij} = \sum_{p=2,6} \left[ \prod_{k\not=i,j}f^p_{ijk}({\bf r}_{ik}
              ,{\bf r}_{jk}) \right] u_p(r_{ij}) O^p_{ij} \ ,
\end{equation}
while the two-body spin-orbit correlation operator is given by
\begin{equation}
     U^{LS}_{ij} = \sum_{p=7,8} u_p(r_{ij}) O^p_{ij} \ ,
\end{equation}
with
\begin{equation}
     O^{p=1,8}_{ij} = [ 1, \sigma_i\cdot\sigma_j, S_{ij},
                      ( {\bf L}\cdot{\bf S} )_{ij} ]\otimes
                      [ 1, \tau_i\cdot\tau_j ] \ .
\label{eq:osix}
\end{equation}
The $U_{ij}$ and $U^{LS}_{ij}$ correlations are induced by the 
two-nucleon interaction.
The equations used to generate the functions $f_c(r_{ij})$ and $u_p(r_{ij})$
are given in Ref.~\cite{W91};
they contain a number of variational parameters to be determined by
minimizing the energy.  
The shape parameters listed in Table~\ref{table:params1} seem to have
negligible $A$ dependence.  
Their values are determined by minimizing the $^{3}$H energy, and are then 
used for all subsequent $A\geq4$ calculations.
There are also a number of parameters that describe the long-range behavior of
the correlation functions which do vary with $A$, as well
as with the Hamiltonian.
Our best values for these parameters are given in Table~\ref{table:params2}.

The $f^c_{ijk}$, $f^p_{ijk}$, and $U_{ijk}$ are three-nucleon correlations 
induced by $v_{ij}$.
The first two have an operator-independent form:
\begin{equation}
     f^{c}_{ijk} = 1 + q^{c}_{1}({\bf r}_{ij}\cdot{\bf r}_{ik})
                                ({\bf r}_{ji}\cdot{\bf r}_{jk})
                                ({\bf r}_{ki}\cdot{\bf r}_{kj})
                                exp(-q^{c}_{2}R_{ijk}) \ ,
\label{eq:fcijk}
\end{equation}
\begin{equation}
     f^{p}_{ijk} = 1 - q^{p}_{1}(1-{\bf \hat r}_{ik}\cdot{\bf \hat r}_{jk})
                                exp(-q^{p}_{2}R_{ijk}) \ ,
\label{eq:fpijkl}
\end{equation}
where $R_{ijk} = r_{ij} + r_{ik} + r_{jk}$.
The $U_{ijk}$ involve additional spin-isospin operators and are somewhat 
more complicated; they are discussed in Ref.~\cite{APW95}.
The $U^{TNI}_{ijk}$ are three-body correlations induced by the three-nucleon
interaction, which we take in the form suggested by perturbation theory:
\begin{equation}
     U^{TNI}_{ijk} = \sum_x \epsilon_x V^x_{ijk}(\tilde{ r}_{ij},
                     \tilde{r}_{jk}, \tilde{ r}_{ki}) \ ,
\label{eq:bestuijk}
\end{equation}
with $\tilde{r}=yr$, $y$ a scaling parameter, and $\epsilon_x$ a strength 
parameter.
Here $x=A$, $C$, and $R$ refers to the different parts of the $N\!N\!N$ 
potential.
With the present Hamiltonian we use the three-body parameters given in 
Table~\ref{table:params3} for all nuclei studied here.

The relative importance and cost of computing different elements of the
full variational wave function $\Psi_V$ are shown in Table~\ref{table:psiv}
for the case of $^6$Li.
The pair wave function $\Psi_P$ is the least expensive to compute, but gives
a rather poor energy.
The successive addition of $U^{TNI}_{ijk}$, $U^{LS}_{ij}$, and $U_{ijk}$
correlations to make up the full $\Psi_V$ lowers the energy by $\sim 2.7$
MeV, but requires $\sim 2.7$ times more computation than $\Psi_P$.
However, about 60\% of the energy gain can be obtained for only a 20\% increase
in computation by using the trial function $\Psi_T$ defined by
\begin{eqnarray}
|\Psi_{T}\rangle = \left[1 + \sum_{i<j<k}\tilde{U}^{TNI}_{ijk}\right] 
|\Psi_P\rangle \ .
\label{eq:psitgfmc}
\end{eqnarray}
Here $\tilde{U}^{TNI}_{ijk}$ is a truncated TNI correlation from which 
the commutator term, $\epsilon_{C}V_{ijk}^{C}$, has been omitted:
\begin{eqnarray}
\tilde{U}^{TNI}_{ijk} = \tilde{\epsilon}_{A}V_{ijk}^{A} +
\epsilon_{R}V_{ijk}^{R} \ .
\end{eqnarray} 
The strength of the anticommutator term is increased to compensate, with
$\tilde{\epsilon}_{A} \sim 1.5\epsilon_{A}$.
From Table~\ref{table:psiv} we see that this simplification gets 90\% of the
gain of adding the full $U^{TNI}_{ijk}$, at $\case{1}{3}$ the cost.
The computer time is reduced so significantly because $\{X^{\pi}_{ij},X^{\pi}_{jk}\}$
can be written as a generalized tensor operator involving the spins of
only nucleons $i$ and $k$; thus the time scales as the number of pairs
rather than the number of triples.
As discussed below, $\Psi_T$ is in fact the most economical starting point 
for the GFMC calculations.

The Jastrow wave function for $A=6$ nuclei is more complicated, as two nucleons 
must be placed in the $p$-shell.  
We use $LS$ coupling to obtain the desired $JM$ value of a given state, 
as suggested in shell-model studies of $p$-shell nuclei~\cite{CK65}.
Different possible $LS$ combinations lead to multiple components in the
Jastrow wave function.
We also allow for the possibility that the central correlations $f_{c}(r_{ij})$ 
and $f^c_{ijk}$ could depend upon the shells ($s$ or $p$) occupied by the
particles and on the $LS$ coupling.
The Jastrow wave function is taken as
\begin{eqnarray}
  |\Psi_J\rangle &=& {\cal A} \left\{
     \left[ \prod_{i<j<k \leq 4}f^{sss}_{ijk} \prod_{l<m \leq 4}f^{ssp}_{lm5}
     f^{ssp}_{lm6} \prod_{n \leq 4}f^{spp}_{n56} \right]\right.  \nonumber\\
  && \left.\left[ \prod_{i<j \leq 4}f_{ss}(r_{ij}) 
     \prod_{k \leq 4} f_{sp}(r_{k5}) f_{sp}(r_{k6}) 
     \sum_{LS} \Big( \beta_{LS} f^{LS}_{pp}(r_{56})
     |\Phi_6(LSJMTT_{3})_{1234:56}\rangle \Big) \right] \right\} \ .
\label{eq:jastrow6}
\end{eqnarray}
The operator ${\cal A}$ indicates an antisymmetric sum over all possible
partitions of the six particles into 4 $s$-shell and 2 $p$-shell ones.
For the two-body correlations we use $f_{ss}(r)=f_c(r)$ from the $^{4}$He 
wave function, while
\begin{eqnarray}
   f_{sp}(r) &=& 
         [a_{sp} + b_{sp}W(r)]f_c(r) + c_{sp}(1-exp[-(r/d_{sp})^2]) \ , \\
   f^{LS}_{pp}(r) &=& 
         [a_{pp} + b_{pp}W(r)]f_c(r) + c^{LS}_{pp}(1-exp[-(r/d_{pp})^2]) \ .
\label{eq:flspp}
\end{eqnarray}
Here we have supplemented the $f_c(r)$ with a long-range tail
and allowed for a short-range modification with a Woods-Saxon factor,
$W(r) = \left\{1+{\rm exp}[(r-R_f)/a_f]\right\}^{-1}$.
The $a_{sp}$, $b_{sp}$, etc., are variational parameters, whose values are 
given in Table~\ref{table:params4}.
For the three-body correlations, our best present trial function has 
$f^{sss}_{ijk} = f^{c}_{ijk}$ as in Eq.(\ref{eq:fcijk}), 
and $f^{ssp}_{ijk} = f^{spp}_{ijk} = 1$.

The $LS$ components of the single-particle wave function are given by:
\begin{eqnarray}
 &&  |\Phi_{6}(LSJMTT_{3})_{1234:56}\rangle = 
     |\Phi_{4}(0 0 0 0)_{1234} 
     \phi^{LS}_{p}(R_{\alpha 5}) \phi^{LS}_{p}(R_{\alpha 6})
     \nonumber \\
 &&  \left\{ [Y_{1m_l}(\Omega_{\alpha 5}) Y_{1m_l'}(\Omega_{\alpha 6})]_{LM_L}
     \times [\chi_{5}(\case{1}{2}m_s) \chi_{6}(\case{1}{2}m_s')]_{SM_S}
     \right\}_{JM} \nonumber \\
 &&  \times [\nu_{5}(\case{1}{2}t_3) \nu_{6}(\case{1}{2}t_3')]_{TT_3}\rangle \ .
\end{eqnarray}
The $\phi^{LS}_{p}(R_{\alpha k})$ are $p$-wave solutions of a particle of
reduced mass $\case{4}{5}m_N$ in an effective $\alpha$-$N$ potential.
They are functions of the distance between the center of mass 
of the $\alpha$ core (which contains particles 1-4 in this partition)
and nucleon $k$, and again may be different for different $LS$ components.
We use a Woods-Saxon potential well:
\begin{equation}
     V_{\alpha N}(r) = V^{LS}_p [1+exp(\frac{r-R_p}{a_p})]^{-1} \ ,
\label{eq:spwell}
\end{equation}
where $V^{LS}_p$, $R_p$, and $a_p$ are variational parameters and we allow 
the depth of the well to vary with the $LS$ composition.
The values of these parameters are also given in Table~\ref{table:params4}.
The wave function is translationally invariant, hence there is no spurious
center of mass motion.

The experimental spectra for $A=6$ nuclei~\cite{AS88} are shown in 
Fig.~\ref{fig:expt6}.
The ground state of $^6$He is strong stable, but decays by $\beta^-$ 
emission with a mean life of 807 ms.
The first excited state is above the threshold for decay to $\alpha+n+n$
and has a width of $\approx110$ keV; we treat it here as a stable state 
with zero width.
In the shell model, the $(J^{\pi};T) = (0^+;1)$ ground state of $^6$He is 
predominantly a $^{2S+1}L[n] = ^1$S[2] state, where we use spectroscopic 
notation to denote the total $L$ and $S$ of the state and the Young 
pattern $[n]$ to indicate the spatial symmetry.
The $(2^+;1)$ first excited state is predominantly a $^1$D[2] state.
We allow for a possible $^3$P[11] admixture in both states, using amplitudes
$\beta_{00}$ and $\beta_{11}$ in the ground state, and $\beta_{20}$ and 
$\beta_{11}$ in the excited state.
After other parameters in the trial function have been optimized, we 
make a series of calculations in which the $\beta_{LS}$ may be different in
the left- and right-hand-side wave functions to obtain the diagonal and
off-diagonal matrix elements of the Hamiltonian and the corresponding 
normalizations and overlaps.
We diagonalize the resulting $2\times2$ energy matrices to find the $\beta_{LS}$
eigenvectors.
The shell model wave functions are orthonormal, but the correlated $\Psi _V$ 
are not.  
Hence the diagonalizations use generalized eigenvalue routines including 
overlap matrices.
We also calculate the position of the three complementary $^3$P[11] states,
with $(J^{\pi};T) = (2^+;1)$, $(1^+;1)$, and $(0^+;1)$; only one of these 
has been tentatively identified experimentally~\cite{MSU96}.
The normalized $\beta_{LS}$ for these different states are given in
Table~\ref{table:beta6}.

The spectrum for $^6$Li contains a $(1^+;0)$ ground state that is 
predominantly $^3$S[2] in character and a triplet of $^3$D[2] excited states
with $(3^+;0)$, $(2^+;0)$, and $(1^+;0)$ components.
In addition, there are $(0^+;1)$ and $(2^+;1)$ excited states that are
the isobaric analogs of the $^6$He ground and excited states.
The $^6$Li ground state is stable, while the $(3^+;0)$ first excited state is
above the threshold for breakup into $\alpha+d$, but is narrow with 
a width of 24 keV.
The $(0^+;1)$ second excited state is even narrower, being unable to decay
to $\alpha+d$ without isospin violation, and thus has a width of only 8 eV.
The higher excitations have widths of 0.5 to 1.7 MeV, but we will treat them
here as well-defined states.
For the $(1^+;0)$ ground and excited states we mix $\beta_{01}$, $\beta_{21}$, 
and $\beta_{10}$ amplitudes by diagonalizing a $3\times3$ energy matrix.
The latter amplitude corresponds to an unobserved $^1$P[11] state, 
which we also obtain in this diagonalization.
However, only the $\beta_{21}$ amplitude contributes to the $(3^+;0)$ and 
$(2^+;0)$ excited states.
Again, Table~\ref{table:beta6} gives a summary of the $\beta_{LS}$ amplitudes.

The ground and first excited states in $^6$Be have the same character 
as those in $^6$He, except that the Coulomb interaction makes the ground 
state a resonance, with a width of 92 keV.
Again, we neglect the resonance character of these states in constructing the
trial function.
Most of the numerical results calculated here for the CIB terms of the 
Hamiltonian are obtained by interchanging neutrons and protons in the 
wave function.
This does not allow for the changes expected in $^6$Be compared
to $^6$He due to the Coulomb force, so we have also made some calculations
adding a Coulomb term $V^C_{\alpha N}(r)$, folded over nuclear form factors, 
to the $V_{\alpha N}(r)$ used to generate the single-particle 
$\phi_{p}(R_{\alpha k})$ functions:
\begin{eqnarray}
  V^C_{\alpha N}(r) = Z\frac{e^2}{r} \left\{\right. 1 &-& \frac{1}{2}exp(-x_{\alpha})
            [2+x_{\alpha}+\frac{4}{1-y^2}] [1-y^{-2}]^{-2}  \nonumber\\
                                               &-& \frac{1}{2}exp(-x_p)
            [2+x_p+\frac{4}{1-y^{-2}}] [1-y^2]^{-2} \left.\right\} \ .
\label{eq:vcan}
\end{eqnarray}
Here $x_{\alpha} = \sqrt{12}r/r_{\alpha}$, $x_p = \sqrt{12}r/r_p$, and
$y=r_{\alpha}/r_p$, with the charge radii $r_{\alpha}=1.65$ fm and 
$r_p = 0.81$ fm.
This additional potential term can be used with strength $Z=0$, 1, or 2 for 
$^6$He, $^6$Li, or $^6$Be, respectively, corresponding to the average Coulomb
interaction between the $\alpha$ core and a $p$-shell nucleon.

The full $A=6$ wave function is constructed by acting on the $|\Psi_J\rangle$,
Eq.(\ref{eq:jastrow6}), with the same $U_{ij}$, $U^{LS}_{ij}$, $U_{ijk}$, 
and $U^{TNI}_{ijk}$ correlations used in $^4$He.
The one exception is that the optimal strength of the $U_{ijk}$ correlations is
reduced slightly as $A$ increases.
In principle, the $U_{ij}$ could be generalized to be different according
to whether particles $i$ and $j$ are in the $s$- or $p$-shell, but this
would require a larger sum over the different partitions and would increase
the computational cost by an order of magnitude.

The Jastrow wave function for $A=7$ nuclei is a straightforward extension
of Eq.(\ref{eq:jastrow6}), with the added specification of the spatial
symmetry $[n]$ of the angular momentum coupling of three $p$-shell nucleons:
\begin{eqnarray}
  |\Psi_{J}\rangle &=& {\cal A} \left\{\right.
     \left[ \prod_{i<j<k \leq 4} f^{sss}_{ijk}
            \prod_{l<m \leq 4}f^{ssp}_{lm5} f^{ssp}_{lm6} f^{ssp}_{lm7}
            \prod_{n \leq 4}f^{spp}_{n56} f^{spp}_{n57} f^{spp}_{n67}
            \ f^{ppp}_{567} \right]  \nonumber\\
                   && \left[
            \prod_{i<j \leq 4}f_{ss}(r_{ij})
            \prod_{k \leq 4} f_{sp}(r_{k5})f_{sp}(r_{k6})f_{sp}(r_{k7})
            \right. \nonumber\\
                   && \left. \sum_{LS[n]}
            \Big( \beta_{LS[n]} \prod_{5 \leq l<m} f^{LS}_{pp}(r_{lm})
            |\Phi_{7}(LS[n]JMTT_{3})_{1234:567}\rangle \Big) \right]
            \left.\right\} \ .
\end{eqnarray}
The single-particle wave function is then
\begin{eqnarray}
 &&  |\Phi_{7}(LS[n]JMTT_{3})_{1234:567}\rangle =
     |\Phi_{4}(0 0 0 0)_{1234} \phi^{LS}_{p}(R_{\alpha 5}) 
     \phi^{LS}_{p}(R_{\alpha 6}) \phi^{LS}_{p}(R_{\alpha 7}) \nonumber \\
 &&  \left\{ [Y_{1m_l}(\Omega_{\alpha 5}) Y_{1m_l'}(\Omega_{\alpha 6})
                     Y_{1m_l^{\prime\prime}}(\Omega_{\alpha 7})]_{LM_L[n]}
     \times [\chi_{5}(\case{1}{2}m_s) \chi_{6}(\case{1}{2}m_s')
             \chi_{7}(\case{1}{2}m_s^{\prime\prime})]_{SM_S}
     \right\}_{JM} \nonumber \\
 &&  \times [\nu_{5}(\case{1}{2}t_3) \nu_{6}(\case{1}{2}t_3')
             \nu_{7}(\case{1}{2}t_3^{\prime\prime})]_{TT_3}\rangle \ .
\end{eqnarray}
There is an implicit complementary symmetry $[n^{\prime}]$ for the spin-isospin
part of the wave function, to preserve the overall antisymmetry,
which we do not show here explicitly; for a detailed discussion of
the symmetry considerations see Appendix 1C of Ref.~\cite{BM69}.
The full wave function is again built up using Eq.(\ref{eq:bestpsiv}) 
with the added definition of the central $p$-shell three-body correlation 
$f^{ppp}_{ijk}=f^{sss}_{ijk}$.

The experimental spectra for $A=7$ nuclei~\cite{AS88} are shown in
Fig.~\ref{fig:expt7}.
In the shell model, the lowest states for the $T=\case{1}{2}$ nuclei $^7$Li 
and $^7$Be have a predominantly $^2$P[3] character, split into 
$(J^{\pi};T)=(\case{3}{2}^-,\case{1}{2})$ ground and
$(\case{1}{2}^-,\case{1}{2})$ first excited states.
These states are all strong stable, the first excited states having 
mean lives $\approx100-200$ fs, while the ground state of $^7$Be decays weakly
with a mean life of 53 d.
Each ground state can also have contributions with a mixed spatial symmetry,
including $^2$P[21], $^4$P[21], $^4$D[21], and $^2$D[21] components, while
the first excited state has admixtures of $^2$P[21], $^4$P[21], $^4$D[21],
and $^2$S[111] amplitudes.
We have diagonalized $5\times 5$ matrices for these states.
Higher in the spectrum is a predominantly $^2$F[3] state that splits 
into $(\case{7}{2}^-,\case{1}{2})$ and $(\case{5}{2}^-,\case{1}{2})$ pieces.
The lower state is mixed with a $^4$D[21] component, while the upper state has 
$^4$P[21], $^4$D[21], and $^2$D[21] contributions.
Again we have performed diagonalizations in the $\beta_{LS[n]}$ amplitudes to
project out the lowest $(\case{7}{2}^-,\case{1}{2})$ and
$(\case{5}{2}^-,\case{1}{2})$ states.

The diagonalizations confirm that the ground and first excited states are
almost pure $^2$P[3] and the second and third excited states are almost
pure $^2$F[3].
We have also calculated the next excited state of each $J$ as given by our
projections to confirm that they lie above these first four states.
The normalized amplitudes of the lowest two states of each $J$ are given in 
Table~\ref{table:beta7}.
These higher excitations include a triplet of predominantly $^4$P[21] states: 
$(\case{5}{2}^-,\case{1}{2})$, $(\case{3}{2}^-,\case{1}{2})$, and
$(\case{1}{2}^-,\case{1}{2})$, and a $(\case{7}{2}^-,\case{1}{2})$ state 
that is predominantly $^4$D[21].
The experimental spectrum shows a similar ordering of states, except that
the $(\case{7}{2}^-,\case{1}{2})$ comes in the midst of the $^4$P[21] states, 
and no second $(\case{1}{2}^-,\case{1}{2})$ state has been identified in this
range of energy excitation.

The spectrum for $T=\case{3}{2}$ states in $A=7$ nuclei is also shown in 
Fig.~\ref{fig:expt7}.
The $(\case{3}{2}^-;\case{3}{2})$ ground state for $^7$He is 0.44 MeV above 
the threshold for breakup into $^6$He+$n$ with a width of 160 keV.
The isobaric analogs have widths of 260, 320, and 1200 keV for $^7$Li, 
$^7$Be, and $^7$B, respectively.
This state can have contributions from $^2$P[21], $^2$D[21], and $^4$S[111]
amplitudes, and we again diagonalize a $3 \times 3$ matrix to evaluate the 
$\beta_{LS[n]}$ components, as shown in Table~\ref{table:beta7}.
We also calculate the first three excited states: 
a $(\case{1}{2}^-;\case{3}{2})$ $^2$P[21] state, 
a $(\case{5}{2}^-;\case{3}{2})$ $^2$D[21] state, and the second mixed
$(\case{3}{2}^-;\case{3}{2})$ state.
None of these excited states have been experimentally identified.

\subsection{Energy Evaluation}

The energy expectation value of Eq.(\ref{eq:expect}) is evaluated using Monte 
Carlo integration.
A detailed technical description of the methods used here can be found in 
Refs.~\cite{W91,CW91,P96}.
Monte Carlo sampling is done both in configuration space and in the
order of operators in the symmetrized product of Eq.(\ref{eq:psip})
by following a Metropolis random walk.
The expectation value for an operator $O$ is given by
\begin{equation}
  \langle O \rangle = \frac 
  {\sum_{p,q} \int d{\bf R} \Psi_{p}^{\dagger}({\bf R}) O \Psi_{q}({\bf R})}
  {\sum_{p,q} \int d{\bf R} \Psi_{p}^{\dagger}({\bf R})   \Psi_{q}({\bf R})} \ .
\label{eq:expecto}
\end{equation}
The subscripts $p$ and $q$ specify the order of operators in the left and 
right hand side wave functions, while the integration runs over the particle 
coordinates ${\bf R}=({\bf r}_1,{\bf r}_2,\ldots,{\bf r}_A)$.
This multidimensional integration is facilitated by introducing a probability
distribution, $W_{pq}({\bf R})$, such that
\begin{equation}
  \langle O \rangle = \frac
  { \sum_{p,q} \int d{\bf R} 
    \left[ \Psi_{p}^{\dagger}({\bf R}) O \Psi_{q}({\bf R}) / 
           W_{pq}({\bf R}) \right] W_{pq}({\bf R}) }
  { \sum_{p,q} \int d{\bf R}
    \left[ \Psi_{p}^{\dagger}({\bf R})   \Psi_{q}({\bf R}) /
           W_{pq}({\bf R}) \right] W_{pq}({\bf R}) } \ .
\end{equation}
This probability distribution is taken to be
\begin{equation}
   W_{pq}({\bf R}) = | {\rm Re}( \langle \Psi_{P,p}^{\dagger}({\bf R})
                                       \Psi_{P,q}({\bf R}) \rangle ) | \ ,
\label{eq:vmc:weight}
\end{equation}
which is constructed from the pair wave function, $\Psi_P$, but with only 
one operator order of the symmetrized product.
This probability distribution is much less expensive to compute than the
full wave function of Eq.(\ref{eq:bestpsiv}) with its spin-orbit and 
operator-dependent three-body correlations, but it typically 
has a norm within 1--2\% of the full wave function.

Expectation values have a statistical error which can be estimated by the
standard deviation $\sigma$:
\begin{equation}
   \sigma = \left[ \frac{ \langle O^2 \rangle - \langle O \rangle ^2}
                        { N-1 } \right] ^{1/2} \ ,
\end{equation}
where $N$ is the number of statistically independent samples.
Block averaging schemes can be used to estimate the autocorrelation
times and determine the statistical error.

The wave function $\Psi$ can be represented by an array of $2^A \times (^A_Z)$
complex numbers, 
\begin{equation}
  \Psi({\bf R}) = \sum_{\alpha} \psi_{\alpha}({\bf R}) |\alpha\rangle \ ,
\label{eq:psivec}
\end{equation}
where the $\psi_{\alpha}({\bf R})$ are the coefficients of each state 
$|\alpha\rangle$ with specific third components of spin and isospin.
This gives arrays with 96, 960, 1280, 2688, and 4480 elements
for $^4$He, $^6$He, $^6$Li, $^7$He, and $^7$Li, respectively.
The spin, isospin, and tensor operators $O^{p=2,6}_{ij}$ contained in
the two-body correlation operator $U_{ij}$, and in the Hamiltonian are
sparse matrices in this basis.
For forces that are largely charge-independent, as is the case here,
we can replace this charge-conserving basis with an isospin-conserving basis
that has $N(A,T) = 2^A \times I(A,T)$ components, where
\begin{equation}
   I(A,T) = \frac{2T+1}{\case{1}{2}A+T+1} 
            \left(\begin{array}{c} A \\ \case{1}{2}A+T \end{array}\right) \ .
\label{eq:numiso}
\end{equation}
This reduces the number of array elements to 32, 576, 320, 1792, and 1792
for the cases given above -- a significant savings.
In practice, the $\tau_i \cdot \tau_j$ operator is more expensive to evaluate
in this basis, but the overall savings in computation are still large.

Expectation values of the kinetic energy and spin-orbit potential require
the computation of first derivatives and diagonal second derivatives of the
wave function.
These are obtained by evaluating the wave function at $6A$ slightly shifted
positions of the coordinates ${\bf R}$ and taking finite differences, 
as discussed in Ref.~\cite{W91}.
Potential terms quadratic in {\bf L} require mixed second derivatives, which
can be obtained by additional wave function evaluations and finite differences.
A rotation trick can be used to reduce the number of additional locations
at which the wave function must be evaluated~\cite{SPF89}.

As a check on the correctness of our Monte Carlo integration, we have evaluated 
the energy expectation value $\langle H \rangle$ for the deuteron using the
exact wave function as input, and match the energy to better than 1 keV.
We have made similar calculations for the triton using a Faddeev wave function,
as discussed in Ref.~\cite{W91}, and obtained agreement with independent
Faddeev calculations at the 10--20 keV level.
For the much more complicated $A=6,7$ wave functions, we also evaluate the 
expectation values $\langle J^2 \rangle$ and $\langle J_z \rangle$ to verify
that they truly have the specified quantum numbers.
A third check is made on the antisymmetry of the Jastrow wave function by
evaluating, at an initial randomized position,
\begin{equation}
   \frac{
    \Psi_J^{\dagger} [ 1 + P^x_{ij}P^{\sigma}_{ij}P^{\tau}_{ij} ] \Psi_J }
  { \Psi_J^{\dagger} \Psi_J } \ , \nonumber
\end{equation}
where $P^{x,\sigma,\tau}_{ij}$ are the space, spin, and isospin 
exchange operators.
This value should be exactly zero for an antisymmetric wave 
function, and it is in fact less than $10^{-9}$ for each pair of particles
in each nuclear state that we study.

A major problem arises in minimizing the variational energy for $p$-shell
nuclei using the above wave functions: there is no variational minimum that
gives reasonable rms radii.
For example, the variational energy for $^6$Li is slightly more bound
than for $^4$He, but is not more bound than for separated $^4$He and $^2$H
nuclei, so the wave function is not stable against breakup into $\alpha + d$
subclusters.
Consequently, the energy can be lowered toward the sum of $^4$He and $^2$H
energies by making the wave function more and more diffuse.
Such a diffuse wave function would not be useful for computing other nuclear 
properties, or as a starting point for the GFMC calculation (see Sec.VI below),
so we constrain our search for optimal variational parameters by requiring the
resulting point proton rms radius, $r_p$, to be close to the experimental 
values for $^6$Li and $^7$Li ground states.
For $^6$He and $^7$Be ground states, and all the excited or resonant states, 
there are no experimental measurements of the charge radii. 
To avoid introducing too many additional parameter values, we construct these
wave functions by making minimal changes to the $^6$Li and $^7$Li wave 
functions, with the added requirement that the excited states should not have 
smaller radii than the ground states.

For $A=6$ nuclei, we begin by selecting parameters to minimize the energy of 
the $^6$Li $^3$S[2] component (the dominant part of the ground state) subject 
to the constraint that $r_p \sim 2.4-2.5$ fm. 
For the other components, only the depth of the single-particle well, 
$V^{LS}_p$, of Eq.(\ref{eq:spwell}), and the tail, $c^{LS}_{pp}$, of the 
$p$-shell pair correlation function of Eq.(\ref{eq:flspp}) are varied, as 
shown in Table~\ref{table:params4}.
The well depth for the $^3$D[2] states is decreased to get the rms radius
of the $(3^+;0)$ excited state larger than the ground state.
The tail is increased for the mixed-symmetry $^1$P[11] state for the same 
reason.
For $^6$He we use the same parameters as in $^6$Li for the corresponding
$^1$S[2], $^1$D[2], and $^3$P[11] states.
The only other difference between $^6$Li and $^6$He wave functions is that
we may turn on the $\alpha$-$N$ Coulomb potential of Eq.(\ref{eq:vcan})
when generating the single-particle radial functions $\phi_p$.
Finally, the diagonalizations are made to determine the $\beta_{LS}$ mixing
coefficients of Table~\ref{table:beta6}.

A similar procedure is followed for the $A=7$ nuclei.
Parameters are selected for the dominant $^2$P[3] state in $^7$Li subject to 
the constraint that $r_p \sim 2.2-2.3$ fm.
The well depth is reduced for the $^2$F[3] states, and the tail is
increased for all the mixed symmetry states.
Afterwards the $\beta_{LS}$ diagonalization is carried out.
Since $^7$Be is a mirror nucleus, it has the same wave function as $^7$Li, 
aside from changing the $\alpha$-$N$ Coulomb potential.
The $^7$He and $^7$B ground states are isobaric analogs to mixed symmetry 
states in $^7$Li, so they use corresponding parameters.

Shell model lore tells us that the lowest state of any given $(J^{\pi};T)$
will be the state with maximal spatial symmetry and smallest $L$ that
can be formed from the allowed couplings, e.g., the $^3$S[2] ground state in
$^6$Li or the $^2$P[3] ground state in $^7$Li.
For the purposes of obtaining a variational upper bound and a GFMC starting
point, we could settle for a $\Psi_V$ constructed using only that $LS[n]$ 
component.
However, by using all the allowed components, we can gain a significant
amount of energy in some cases and, as is be discussed below, this gain 
persists in our GFMC propagations.
For the $A=6$ nuclei, the diagonalizations for the $(0^+;1)$, $(1^+;0)$, and
$(2^+;1)$ states improve the lower state by 0.25 to 0.5 MeV.
In the first four $^7$Li $T=\case{1}{2}$ states, the mixing is much less, 
and improvements are at most 0.15 MeV.
However, for $^7$He, there is a gain of 0.75 MeV, probably because there
are two states of identical symmetry that only differ by 1 in $L$.

The diagonalizations have the additional benefit that we can predict where 
the next higher excited state of each $(J^{\pi};T)$ lies.
This allows us to confirm that the Hamiltonian is not predicting any unusually 
low excitations that are not observed experimentally; e.g., the second 
$(\case{3}{2}^-;\case{1}{2})$ and $(\case{1}{2}^-;\case{1}{2})$ states 
in $^7$Li do not appear below the first $(\case{7}{2}^-;\case{1}{2})$ and 
$(\case{5}{2}^-;\case{1}{2})$ states.

\section{GREEN'S FUNCTION MONTE CARLO}

The aim of the GFMC method is to project out the exact lowest energy
state, $\Psi_{0}$, associated with a chosen set of quantum numbers, 
from an approximation $\Psi_{T}$ to that state.  
The method used here is essentially identical to that used previously 
to calculate nuclei with A$\leq$ 6~\cite{PPCW95},
with the exception that we have now incorporated the exact
two-body propagator in the imaginary-time propagation.  In this section
we describe the algorithm in some detail, in particular relating it
to algorithms commonly used for scalar interactions.
For simplicity of notation we will not
make the distinction between $H^{\prime}$ and $H$ (and their respective
components) that was introduced with Eq.(\ref{eq:hgfmc}); the reader
will want to remember that we in fact use the simpler $H^{\prime}$ in
our GFMC propagator.

GFMC projects out the lowest energy ground state using 
$\Psi_0 = \lim_{\tau \rightarrow \infty} \exp [ - ( H - E_0) \tau ] \Psi_T$.
The eigenvalue $E_{0}$ is calculated exactly while other expectation values
are generally calculated neglecting terms of order $|\Psi_{0}-\Psi_{T}|^{2}$ 
and higher.  
In contrast, the error in the variational energy, $E_{V}$, is of order 
$|\Psi_{0}-\Psi_{T}|^{2}$, and other expectation values calculated with 
$\Psi_{T}$ have errors of order $|\Psi_{0}-\Psi_{T}|$.

We use the $\Psi_{T}$ of Eq.(\ref{eq:psitgfmc}) as our initial trial function
and define the propagated wave function $\Psi(\tau)$ as
\begin{eqnarray}
   \Psi(\tau) = e^{-({H}-E_{0})\tau} \Psi_{T} ;
\end{eqnarray}
obviously $\Psi(\tau=0) =  \Psi_{T}$ and
$\Psi(\tau \rightarrow \infty) = \Psi_{0}$.
Introducing a small time step, $\triangle\tau$, $\tau=n\triangle\tau$, gives
\begin{eqnarray}
\Psi(\tau) = \left[e^{-({H}-E_{0})\triangle\tau}\right]^{n} \Psi_{T}.
\end{eqnarray}
The $\Psi (\tau)$ is represented by a vector function of $\bf R$ using
Eq.(\ref{eq:psivec}), and the
Green's function, $G_{\alpha\beta}({\bf R},{\bf R}^{\prime})$ is a matrix
function of $\bf R$ and ${\bf R}^{\prime}$ 
in spin-isospin space, defined as
\begin{eqnarray}
G_{\alpha\beta}({\bf R},{\bf R}^{\prime})= \langle {\bf 
R},\alpha|e^{-({H}-E_{0})\triangle\tau}|{\bf R}^{\prime},\beta\rangle.
\label{eq:gfunction}
\end{eqnarray}
It is calculated with leading errors of order $(\triangle\tau)^{3}$ as
discussed in Sec.~IV.2.  The errors in the full calculation are
determined by the difference between the (artificial) Hamiltonian for which
the propagator is exact and H.  This difference is of order
$(\triangle \tau)^2$, and $\triangle\tau$ is 
chosen to be small enough that this total error is negligible.  
Omitting
spin-isospin indices for brevity, $\Psi({\bf R}_{n},\tau)$ is given by
\begin{eqnarray}
\Psi({\bf R}_{n},\tau) = \int G({\bf R}_{n},{\bf R}_{n-1})\cdots G({\bf 
R}_{1},{\bf R}_{0})\Psi_{T}({\bf R}_{0})d{\bf R}_{n-1}\cdots{\bf R}_{1}d{\bf 
R}_{0}.
\label{eq:gfmcpsi}
\end{eqnarray}

The mixed expectation value of an operator $O$ is defined as:
\begin{eqnarray}
\langle O \rangle_{Mixed} & = & \frac{\langle \Psi_{T} | O | 
\Psi(\tau)\rangle}{\langle \Psi_{T} | \Psi(\tau)\rangle} \nonumber \\
& = & \frac{ \int d {\bf P}_n 
\Psi_{T}^{\dagger}({\bf R}_{n}) O G({\bf R}_{n},{\bf 
R}_{n-1})\cdots G({\bf R}_{1},{\bf R}_{0})\Psi_{T}({\bf R}_{0})}
{\int d{\bf P}_{n} \Psi_{T}^{\dagger}
({\bf R}_{n})G({\bf R}_{n},{\bf R}_{n-1}) \cdots G({\bf 
R}_{1},{\bf R}_{0})\Psi_{T}({\bf R}_{0})}~, \nonumber \\
\label{eq:expectation}
\end{eqnarray}
where ${\bf P}_{n} = {\bf R}_{0},{\bf R}_{1},\cdots,{\bf R}_{n}$ denotes the 
`path', and
\begin{eqnarray}
d{\bf P}_{n} = d{\bf R}_{0} d{\bf R}_{1}\cdots d{\bf R}_{n}~.
\end{eqnarray}
In GFMC, the integral over the paths is carried out stochastically.  
Generally,
the required expectation values are calculated approximately
from the variational $\Psi_T$ and mixed expectation values.  Let
\begin{eqnarray}
\Psi(\tau) = \Psi_{T} + \delta\Psi(\tau).
\end{eqnarray}
Retaining only the terms of order $\delta\Psi(\tau)$, we obtain
\begin{eqnarray}
\langle O (\tau)\rangle & = & 
\frac{\langle\Psi(\tau)| O |\Psi(\tau)\rangle}{\langle\Psi(\tau)|\Psi(\tau)\
\rangle} \nonumber ~,\\
& \approx & \langle O (\tau)\rangle_{Mixed} + [\langle O (\tau)\rangle_{Mixed} 
- \langle O \rangle_T] ~,
\label{eq:pc_gfmc}
\end{eqnarray}
where
\begin{equation}
\langle O \rangle_T = 
\frac{\langle\Psi_{T}| O |\Psi_{T}\rangle}{\langle\Psi_{T}|\Psi_{T}\rangle} ~.\\
\end{equation}
More accurate evaluations of $\langle O (\tau)\rangle$ are possible,
\cite{Kalos67}
essentially by measuring the observable at the mid-point of the path.
However, such estimates require a propagation 
twice as long as the mixed estimate.
Since we are limited in the present calculations to a total propagation
time of $0.06 $MeV$^{-1}$, we use the approximation~(\ref{eq:pc_gfmc}).

An important exception to the above is the energy, $E_{0}$ given by 
$\langle{H}(\tau\rightarrow\infty)\rangle$.  The 
$\langle{H}(\tau)\rangle_{Mixed}$ can be re-expressed as
(Ref. \cite{CepKal79})
\begin{eqnarray}
\langle{H}(\tau)\rangle_{Mixed} = \frac{\langle \Psi_{T} |  
e^{-({H}-E_{0})\tau /2}{H} e^{-({H}-E_{0})\tau /2} | 
\Psi_{T}\rangle}{\langle \Psi_{T} |e^{-({H}-E_{0})\tau /2} 
e^{-({H}-E_{0})\tau /2}| \Psi_{T}\rangle} ~,
\end{eqnarray}
since the propagator $\exp [ - (H - E_0) \tau ] $ commutes with the
Hamiltonian.
Thus $\langle{H}(\tau)\rangle_{Mixed}$ approaches $E_{0}$ in the limit 
$\tau\rightarrow\infty$, and furthermore, being an expectation value of
${H}$, it obeys the variational principle
\begin{eqnarray}
\langle{H}(\tau)\rangle_{Mixed} \geq E_{0}~.
\end{eqnarray}
If a simpler $H^{\prime}$ is being used to construct the GFMC propagator,
then these equations apply to $\langle{H}^{\prime}(\tau)\rangle$,
and $\langle(H-H^{\prime})\rangle$ must be evaluated using 
Eq.(\ref{eq:pc_gfmc}).

Given these expressions, two basic elements are required for any GFMC
calculation.  The first element is the choice of short-time propagator
$\exp[ - ( H - E_0 ) \Delta \tau]$ and the second is a method for sampling
the paths.  We discuss each of these elements in turn.

\subsection{The Short-Time Propagator}

The short-time propagator should allow as large a time step $\triangle\tau$ as 
possible, since the total computational time for propagation is proportional to 
$1/\triangle\tau$.  Earlier calculations \cite{PPCW95,C87,C88} used the propagator obtained 
from the Feynman formulae.  Ignoring three-nucleon interaction terms in 
${H}$, it is given by
\begin{eqnarray}
e^{-{H}\triangle\tau} = \left[{\cal 
S}\prod_{i<j}e^{-v_{ij}\triangle\tau/2}\right] e^{-K\triangle\tau}\left[{\cal 
S}\prod_{i<j}e^{-v_{ij}\triangle\tau/2}\right] + {\cal O}(\triangle\tau^{3})~.
\label{eq:propagator1}
\end{eqnarray}
Note that it is useful to symmetrize the product of 
$e^{-v_{ij}\triangle\tau/2}$ when $\left[v_{ij},v_{jk}\right] \neq 0$, 
in order to reduce the error per iteration.  The nuclear 
$v_{ij}$ has a repulsive core of order GeV.  The main error in the above 
propagator comes from terms in $e^{-{H}\triangle\tau}$ having 
multiple $v_{ij}$, like $v_{ij}Tv_{ij}(\triangle\tau)^{3}$ for example, which
can become large when particles $i$ and $j$ are very close.  In order to make
them negligible a rather small $\triangle\tau \sim 0.1$ GeV$^{-1}$ is used
with the above propagator.  
The matrix elements of the propagator are given by:
\begin{eqnarray}
G_{\alpha\beta}({\bf R},{\bf R}^{\prime})= G_{0}({\bf R},{\bf 
R}^{\prime})\langle\alpha|\left[{\cal S}\prod_{i<j}e^{-v_{ij}({\bf 
r}_{ij})\triangle\tau/2}\right]\left[{\cal S}\prod_{i<j}e^{-v_{ij}({\bf 
r}^{\prime}_{ij})\triangle\tau/2}\right]|\beta\rangle~,
\end{eqnarray}
\begin{eqnarray}
G_{o}({\bf R},{\bf R}^{\prime}) = \langle {\bf R}|e^{-{K}\triangle\tau}|{\bf 
R}^{\prime}\rangle =  \left[ \sqrt{\frac{m}
{2\pi\hbar^{2} \triangle\tau}}\, \right]^{3A}\exp\left[\frac{-({\bf R}-{\bf 
R}^{\prime})^2}{2\hbar^{2}\triangle\tau/m}\right]~.
\end{eqnarray}

However, it is well known from the studies bulk helium atoms
\cite{CepRMP95}
that including the exact two-body propagator allows much larger
time steps. This short-time propagator is
\begin{eqnarray}
G_{\alpha\beta}({\bf R},{\bf R}^{\prime})= G_{0}({\bf R},{\bf 
R}^{\prime})\langle\alpha|\left[{\cal S}\prod_{i<j}\frac{g_{ij}({\bf 
r}_{ij},{\bf r}_{ij}^{\prime})}{g_{0,ij}({\bf r}_{ij},{\bf r}_{ij}^{\prime})} 
\right] |\beta\rangle~,
\label{eq:propagator2}
\end{eqnarray}
where $g_{ij}$ is the exact two-body propagator,
\begin{eqnarray}
g_{ij}({\bf r}_{ij},{\bf r}_{ij}^{\prime}) = \langle{\bf 
r}_{ij}|e^{-H_{ij}\triangle\tau}|{\bf r}_{ij}^{\prime}\rangle~,
\end{eqnarray}
\begin{eqnarray}
{H}_{ij} = -\frac{\hbar^{2}}{m}\nabla^{2}_{ij} + v_{ij}~,
\end{eqnarray}
and $g_{0,ij}$ is the free two-body propagator,
\begin{eqnarray}
g_{0,ij}({\bf r}_{ij},{\bf r}_{ij}^{\prime}) = \left[ 
\sqrt{\frac{\mu}{2\pi\hbar^{2} \triangle\tau}}\, 
\right]^3 \exp\left[-\frac{({\bf r}_{ij}-{\bf 
r}_{ij}^{\prime})^{2}}{2\hbar^{2}\triangle\tau/\mu}\right]~,
\label{eq:gaussian}
\end{eqnarray}
where $\mu = m/2$ is the reduced mass. All terms containing any
number of the same $v_{ij}$ and $K$ are treated exactly in this propagator, 
as we have included the imaginary-time equivalent of the full two-body
scattering amplitude.
It still has errors of order $(\triangle\tau)^{3}$, however they are from
commutators of 
terms like $v_{ij}Tv_{ik}(\triangle\tau)^{3}$ which become large only when
both pairs $ij$ and $ik$ are close.  Since this is a rare occurrence, a five
times larger time step $\triangle\tau \sim 0.5$ GeV$^{-1}$ can be used
for the present studies of light nuclei.  In the case of bound states of
helium atoms a $\sim 30$ times larger time step can be used with the
propagator (\ref{eq:propagator2}) than with (\ref{eq:propagator1}) presumably
because the inter-atomic potentials have a relatively harder core, and they 
commute with each other.

Finally, including the three-body forces and the $E_{0}$ in 
Eq.(\ref{eq:gfunction}), the complete propagator is given by
\begin{eqnarray}
G_{\alpha\beta}({\bf R},{\bf R}^{\prime})& = &e^{E_{o}\triangle\tau}G_{0}({\bf 
R},{\bf R}^{\prime})\exp[{- \sum (V^{R}_{ijk}({\bf R})+ V^{R}_{ijk}({\bf 
R^{\prime}}))\frac{\triangle\tau}{2}}] \nonumber \\ 
& & \langle\alpha|I_{3}({\bf R})|\gamma\rangle\langle\gamma|\left[{\cal 
S}\prod_{i<j}\frac{g_{ij}({\bf r}_{ij},{\bf r}_{ij}^{\prime})}{g_{0,ij}({\bf 
r}_{ij},{\bf r}_{ij}^{\prime})} \right] |\delta\rangle\langle\delta|I_{3}({\bf 
R}^{\prime})|\beta\rangle~,
\label{eq:fullprop}
\end{eqnarray}
\begin{eqnarray}
I_{3}({\bf R}) = \left[1 - \frac{\triangle\tau}{2}\sum V^{2\pi}_{ijk}({\bf 
R})\right]~.
\end{eqnarray}
The exponential of $V^{2\pi}_{ijk}$ is expanded to first order in 
$\triangle\tau$ thus, there are additional error terms of the form 
$V^{2\pi}_{ijk}V^{2\pi}_{i'j'k'}(\triangle\tau)^{2}$.  However, they have 
negligible effect since $V^{2\pi}_{ijk}$ has a magnitude of only a few MeV.
It was verified that the results for $^{4}$He do not show any change, outside
of statistical errors, when $\triangle\tau$ is decreased from 0.5 GeV$^{-1}$.

\subsection{Calculation of $g_{ij}$}

The pair propagator $g_{ij}$ is a matrix in the two-body spin-isospin space, 
and obeys the equation
\begin{eqnarray}
[ \frac{\partial}{\partial \tau} + H_{ij} ]\  g_{ij} 
({\bf r},{\bf r^{\prime}}; \tau) = 0~.
\end{eqnarray}
As the Hamiltonian naturally decomposes into eigenstates of the two-body
spin and isospin, so does the propagator $g_{ij}$. In addition, it
obeys the convolution equation
\begin{eqnarray}
g_{ij}({\bf r},{\bf r}^{\prime};\tau+\tau^{\prime}) = \int d^{3}{\bf 
r}^{\prime\prime}g_{ij}({\bf r},{\bf r}^{\prime\prime};\tau)g_{ij}({\bf 
r}^{\prime\prime},{\bf r}^{\prime};\tau^{\prime})~,
\label{eq:convolution}
\end{eqnarray}
with the initial condition
\begin{equation}
g_{ij} ({\bf r},{\bf r^{\prime}}; \tau = 0 ) = \delta ( {\bf r},{\bf 
r^{\prime}})~.
\end{equation}

To calculate $g_{ij}$, we use the techniques developed by Schmidt and
Lee\cite{SL95} for scalar interactions.  The basic idea is to
use the convolution equation, Eq.(\ref{eq:convolution}), to
write $g_{ij}$ as a product over N steps:
\begin{equation}
g_{ij}({\bf r}_N,{\bf r}_0; \triangle \tau ) = 
\prod_{i=1}^{N}
g_{ij}({\bf r}_{i},{\bf r}_{i-1}; \epsilon),
\end{equation}
with $\epsilon = \triangle \tau / N$
and an implied integration over intermediate points.
If we use a symmetric expression for the short-time propagator
$g_{ij}(\epsilon)$ such as
\begin{equation}
g_{ij}({\bf r},{\bf r}^{\prime};\epsilon) = e^{-v_{ij}({\bf 
r})\epsilon/2}g_{0,ij}({\bf r},{\bf r}^{\prime};\epsilon)e^{-v_{ij}({\bf 
r}^{\prime})\epsilon/2}~, \\
\end{equation}
the errors for $g_{ij} (\triangle \tau)$ contain only even powers of $1/N$,
starting with $1/N^2$.\cite{SL95}  By evaluating $g_{ij}({\bf r},{\bf 
r}^{\prime};\triangle\tau)$
for several values of N (and consequently $\epsilon$) and extrapolating to 
$\epsilon\rightarrow 0$
the $g_{ij}$ can be calculated with high ($\sim$ 10 digit) accuracy.

Evaluation of $g_{ij}$ can be carried out in various ways. We have chosen to 
expand the propagator in partial waves denoted by $J M, T T_{z}, S$, 
and $L$,
thus replacing the three-dimensional integral in Eq.(\ref{eq:convolution})
with many one-dimensional integrals.  The two-nucleon interaction $v_{ij}$ 
has a simple form, $v_{JTS}^{LL^{\prime}}(r)$, in these partial waves.  The
interaction is diagonal ($L=L^{\prime}$) in $S=0$ and $1$ waves with $L=J$,
and it couples the $S=1$, $L$, $L^{\prime}$ = $J \pm 1$ waves.

The $g_{ij}({\bf r},{\bf r}^{\prime};\triangle\tau)$ is 
written as a sum over partial waves,
\begin{eqnarray}
g_{ij}({\bf r},{\bf r}^{\prime};\triangle\tau) = 
\sum_{JM}\sum_{TT_{z}}
\sum_{SLL^{\prime}}
\chi_{T,T_{z}}
{\cal Y}^{M}_{JLS}(\hat{{\bf r}})
\frac{g_{JTS}^{LL^{\prime}}(r,r^{\prime};\triangle\tau) }
{r r^{\prime}}
{\cal Y}^{\dagger M}_{JL^{\prime}S}
(\hat{{\bf r}^{\prime}})\chi^{\dagger}_{T,T_{z}}~,
\label{eq:partialw}
\end{eqnarray}
where $\chi_{T,T_{z}}$ denote isospin states, and ${\cal Y}^{M}_{JLS}
(\hat{{\bf r}})$ are standard spin-angle functions that depend upon spins 
and the directions $\hat{{\bf r}}$ and $\hat{{\bf r}^{\prime}}$.  
In uncoupled channels the partial wave 
propagators $g_{JTS}^{LL^{\prime}}(r,r^{\prime};\triangle\tau)$ are scalar
functions of the magnitudes $r$ and $r^{\prime}$, while for coupled channels
they are 2x2 matrix functions with $L,L^{\prime} = J {\pm} 1$.
The partial wave propagators
obey the one-dimensional convolution equation
\begin{eqnarray}
g_{JTS}^{LL^{\prime}}(r,r^{\prime};\tau+\epsilon) & = & 
\int dr^{\prime\prime} 
\exp[-v_{JTS}^{LL^{\prime\prime}}(r)\frac{\epsilon}{2}] 
g_{0,ij}^{L^{\prime\prime}}(r,r^{\prime\prime};\epsilon)
\exp[-v_{JTS}^{L^{\prime\prime}L^{\prime\prime\prime}}
(r^{\prime\prime})\frac{\epsilon}{2}] \nonumber \\
& & 
g_{JTS}^{L^{\prime\prime\prime}L^{\prime}}(r^{\prime\prime},r^{\prime};\tau),
\label{eq:partconv}
\end{eqnarray}
where $g_{0,ij}^{L}$ is the free propagator in partial waves with angular   
momentum $L$,
\begin{equation}
g_{0,ij}^L =  \frac{x}{\sqrt{r r^{\prime}}}  I_L(x),
\end{equation}
with
\begin{equation}
x \equiv \frac { r r^{\prime} \mu } { 2 \hbar^2 \tau }.
\end{equation}
These expressions are used to obtain $g_{JTS}^{LL^{\prime}}
(r,r^{\prime};\tau)$ for several values of $\epsilon$, 
extrapolating to $\epsilon~\!\!\rightarrow~\!\!0$ until a specific
error tolerance has been reached. It is important to obtain these
`channel' propagators very accurately, because they must be summed
to reproduce a Gaussian falloff in the angular variables that
is present in the full propagator.
Fast Fourier
transforms are used to switch between momentum and coordinate space, where
the kinetic and potential terms, respectively, are diagonal and can be 
trivially exponentiated.

We then sum over partial waves to obtain the full two-body propagator.
If we were to include only the (physical) anti-symmetric two-body channels,
the complete two-body propagator would also be anti-symmetric, and hence
for small $\tau$ the propagator would have two peaks, 
one near the original point and another (with a minus sign
in symmetric spin-isospin states), 
near the point corresponding to the interchange of the particles.
In principle we could use this propagator by sampling paths with an
arbitrary permutation at each step, and perhaps cancel some 
noise arising from unphysical symmetric states.  
However the propagation distance is governed by the Gaussian behavior
of $G_0$, which is much shorter-ranged than the average pair 
separation. Hence, any cancellation would be very small.
Instead, we simply use the Argonne $v^{\prime}_8$ potential in unphysical
states, and include  them in the propagator.
In essence this corresponds to treating the particles as Boltzmann particles
for purposes of the propagator.  Since one always computes overlaps
with completely anti-symmetric states, this is perfectly acceptable.

One also has complete freedom to choose
an arbitrary interaction in the unphysical channels,
but the present choice retains the property of a positive definite Green's
function in spin-singlet channels.   The propagator in
$S=0$ states is positive definite, since it is for small $\epsilon$
and the convolution form of the Eq.(\ref{eq:convolution}) preserves this 
property.
The choice of the Boltzmann propagator allows us to simply sample
Gaussians centered on the identity permutation when choosing the paths.

It is also important to include many partial waves in the calculation.
The starting $g_{ij}(r,r^{\prime};\epsilon)$ is a narrow Gaussian of 
width $\sqrt{(4\hbar^{2}\epsilon/m)}$ which is $< 0.1$ fm for 
$\epsilon < 0.1$ GeV$^{-1}$.   Hence
a large number of partial waves are required
to reproduce it accurately.  
The propagator in all $J\leq55$ states is calculated
from Eq.(\ref{eq:partconv}); beyond this we use 
simple approximations including the analytically known
propagators for free particles.  Keeping a much smaller set of partial
waves would yield the same answer in an exact quadrature, however
it can dramatically increase the statistical error in a Monte Carlo
calculation. For example, the positive definite property described above
is recovered numerically only for large numbers of partial waves.

The terms having $L^2$ and $(L\cdot S)^2$ operators are not presently included
in the propagator.  
These terms, like others that depend quadratically
on the relative momentum between the interacting particles, represent
changes in the mass of the particles due to interactions.  They can in
principle be included by using appropriate effective masses in the
kinetic energy propagator $g_{0,ij}$~\cite{C90}.
Unfortunately, there is a strong
spin-isospin dependence in the $L^2$ and $(L\cdot S)^2$ interactions, which
then makes the $g_{0,ij}$ spin-isospin dependent.  Attempts to use
them have generally led to large statistical errors.
For this reason the propagator uses the approximate
$v^{\prime}_8$ interaction operator.

Calculating this propagator is computationally intensive.  Therefore,
prior to the GFMC calculation, we sum the propagator over partial waves
and store the full sum on a grid.  Storing the 
$g_{JTS}^{LL^{\prime}}(r,r^{\prime};\triangle\tau)$ is impractical, both
because of memory requirements and the fact that summing over waves for
each ${\bf r}$ and ${\bf r}^{\prime}$ would be computationally expensive.
For a spin-independent interaction, the propagator $g_{ij}$ would depend
only upon the two magnitudes $r$ and $r^{\prime}$ and the angle
$cos (\theta) =  \hat{\bf r} \cdot \hat{\bf r}^{\prime}$ between
them.  
Here, though, there is also a dependence upon the spin quantization axis.
Rotational symmetry allows one 
to calculate the spin-isospin components of
$g_{ij}({\bf r},{\bf r}^{\prime})$ for 
any ${\bf r}$ and ${\bf r}^{\prime}$ by simple SU3 
spin rotations and values of $g_{ij}$
on a grid of initial points ${\bf r}=(0,0,z)$ and final points 
${\bf r}^{\prime}=(x^{\prime},0,z^{\prime})$.  
The $x-z$ plane is chosen because
the ${\cal Y}^{M}_{JLS}$ are real there.  In addition, the fact that
the propagator is Hermitian allows us to store only the values for
$z > z^{\prime}$.  In the $z$-direction we take an evenly spaced grid
of 0.02 fm extending up to 6 fm.  Beyond 6 fm the $v_{ij}$ is weak and it
is sufficient to use Eq.(\ref{eq:propagator2}) to calculate $g_{ij}$.
The propagator falls off approximately as a Gaussian, Eq.(\ref{eq:gaussian}),
with range parameter
$2\hbar\sqrt{\triangle\tau/m} \sim 0.3$ fm for $\triangle\tau$ = 0.5 
GeV$^{-1}$.  Thus the $x^{\prime}$ and $z-z^{\prime}$ grids have maximum
values of $\sim 0.9$ fm and are nonuniform.

\subsection{Sampling of the Paths}

The remaining task in a GFMC calculation is to sample a set of
paths; in order to maintain a reasonable statistical error we sum
explicitly over all spin-isospin states of the system for each
path.  To choose the paths we follow as closely as possible 
the standard practice for scalar interactions, as we
have done in previous work~\cite{PPCW95,C88,C87,C90}.  In this section 
we compare the standard method with that used for nuclear systems.

The integrals in Eq.(\ref{eq:expectation}) for 
$\langle\hat{O}(\tau)\rangle_{Mixed}$
are carried out stochastically using a relative probability function 
$P({\bf P})$ to sample the paths.  
Each path consists of a set of n steps, where each step contains a sample of 
3A particle coordinates, as well as sets of operator orders used to sample 
the symmetrization operators ${\cal S}$ for the pair operators in the trial 
wave function, Eq.(\ref{eq:psitgfmc}), and the propagator,
Eq.(\ref{eq:fullprop}).
The ensemble of the sampled paths is denoted by $\{{\bf P}\}$; and
contains $N_{p}$ paths.
For each path ${\bf P}$ we define
\begin{eqnarray}
N_{{\bf P}} = \Psi_{T}^{\dagger}({\bf R}_{n})\hat{O}G({\bf R}_{n},{\bf 
R}_{n-1})\cdots G({\bf R}_{1},{\bf R}_{0})\Psi_{T}({\bf R}_{0})/P({\bf P})~, \\
D_{{\bf P}} = \Psi_{T}^{\dagger}({\bf R}_{n})G({\bf R}_{n},{\bf R}_{n-1})\cdots 
G({\bf R}_{1},{\bf R}_{0})\Psi_{T}({\bf R}_{0})/P({\bf P}) ~,
\end{eqnarray}
and
\begin{eqnarray}
\langle\hat{O}(\tau)\rangle_{Mixed} = (\sum_{\{{\bf P}\}} N_{{\bf 
P}})/(\sum_{\{{\bf P}\}} D_{{\bf P}})~,
\end{eqnarray}
with a statistical error determined by the correlated variance of 
$N_{{\bf P}}$ and $D_{{\bf P}}$
and proportional to $1/\sqrt{N_{p}}$.
The relative probability function $P({\bf P})$, should be
chosen to minimize this statistical error.

Different schemes for sampling the paths with probability $P({\bf P})$
are possible.
In finite-temperature simulations, one typically retains the entire
history of the path and uses a Metropolis scheme to sample them~\cite{CepRMP95}.
For zero-temperature simulations, however, it is generally more efficient to
sample the paths through a branching random walk.
Points along the path are generated iteratively
through an importance-sampling procedure.
Only the amplitudes of the propagated wave function and the
accumulated weight of the path need be retained for each configuration.
We discuss the algorithm used for nuclear spin-isospin dependent
interactions after first describing the algorithm for spin-independent
interactions.

For scalar interactions and real Hamiltonians,
the particles can be assigned specific spin states in $\Psi_{T}({\bf R})$ 
which never change during propagation, $G({\bf R},{\bf R}^{\prime})$ is
a real, positive function with finite norm, and $\Psi_{T}$, and
consequently $\Psi(\tau)$, can be chosen as a real scalar function.  The 
$P({\bf P})$ is commonly taken as
\begin{eqnarray}
P({\bf P}) = \prod_{i=1,n} \left[ I({\bf R}_{i}) G({\bf R}_{i},{\bf 
R}_{i-1}) \frac{1}{I({\bf R}_{i-1})}\right] 
I({\bf R}_0) |\Psi_{T}({\bf R}_{0})|~.
\label{eq:scalweight}
\end{eqnarray}
The importance function $I({\bf R})$ is used in sampling and hence
should be positive definite, it is often taken to be the magnitude
of the trial wave function,
\begin{equation}
I({\bf R})  = | \Psi_T ({\bf R})|~.
\label{eq:importancefn}
\end{equation}
The initial configurations are sampled from 
$I({\bf R}_0)  | \Psi_T ({\bf R}_0)|$.
The quantity in brackets in Eq.(\ref{eq:scalweight}) is referred
to as the importance-sampled Green's function $G_I$,
\begin{equation}
G_I ({\bf R}_{i}, {\bf R}_{i-1} ) =
\left[ I({\bf R}_{i}) G({\bf R}_{i},{\bf
R}_{i-1}) \frac{1}{I({\bf R}_{i-1})}\right]~.
\label{eq:importancegf}
\end{equation}
The probability of the path P({\bf P}) depends implicitly
upon all of the steps in the path, but is decomposed into
an initial weight 
$I({\bf R}_0) |\Psi_{T}({\bf R}_{0})|$,
times a product of weights for each step.

Using this $P(\bf P)$
in our expressions for $N_{{\bf P}}$ and $D_{{\bf P}}$ we 
arrive at
\begin{equation}
N_{{\bf P}}  =  
\frac{\Psi_{T}({\bf R}_{n})\hat{O}}
{I({\bf R}_{n})}
\frac{\Psi_{T}({\bf R}_{0})}
{|\Psi_{T}({\bf R}_{0})|}~, 
\label{eq:numerator}
\end{equation}
\begin{eqnarray}
D_{{\bf P}} = 
\frac{\Psi_{T}({\bf R}_{n})}{I({\bf R}_{n})}
\frac{\Psi_{T}({\bf R}_{0})}{|\Psi_{T}({\bf R}_{0})|}~. 
\label{eq:denominator}
\end{eqnarray}
In the ideal case of a Bose ground state, $\Psi_{T}({\bf R})$ is positive
for all ${\bf R}$; choosing $I({\bf R}) = \Psi_T ({\bf R}) $
yields $D_{{\bf P}} = 1$ with zero variance and the variance of 
$N_{{\bf P}}$ is acceptable for many interesting operators.  
In particular, if 
$\Psi_{T}({\bf R})$ is close to the ground state of ${H}$, then
\begin{eqnarray}
\frac{\Psi_{T}({\bf R}){H}}{\Psi_{T}({\bf R})} \sim E_{0}~,
\end{eqnarray}
and the $N_{{\bf P}}$ for $\langle{H}(\tau)\rangle_{Mixed}$ will have a 
small
variance.  Many properties of Bose liquid and solid $^{4}$He and its
drops have been studied with GFMC \cite{CepRMP95,PZPWH83,PWP85}
using this probability density.

In contrast, the wave functions of simple Fermi systems have domains of 
positive and negative signs separated by nodal surfaces.  The importance
function $I$ given in Eq.(\ref{eq:importancefn})
must be increased slightly near the nodal
surfaces to allow diffusion between the domains.
When the 
path ${\bf P}$ crosses a nodal surface its $D_{{\bf P}}$ and $N_{{\bf P}}$ 
change
sign.  At small $\tau$, few paths are long enough to cross nodal surfaces
and the variance is small.  As $\tau$ increases, many paths cross nodal
surfaces, the variance increases and the average value of $D_{{\bf P}}$ 
decreases.
This problem is called `the Fermion sign problem', and it limits the maximum
value of $\tau$ up to which the state can be propagated.\cite{SK84,LGSWSS90}  
Generally the calculations are continued until the statistical
error increases beyond an `acceptable' point; we 
iterate until $\tau = 0.06 {\rm MeV}^{-1}$ for most nuclei studied here.

For local spin-isospin independent interactions, $^{3}$H,
$^{3}$He, and $^{4}$He nuclei would be completely spatially symmetric
and no sign problem would exist.  For more realistic local interactions, the
dominant spatially-symmetric component of the wave function and the relatively
large excitation energies 
imply that the sign problem is not very significant
for three- and four-body nuclei~\cite{CFP88}.
However, it does limit the propagation of states 
with $A > 4$ which must have nodal surfaces as required by antisymmetry,
and also all calculations with non-local interactions.

Implementing the algorithm  to sample the paths
is straightforward.
Choosing $I({\bf R}) = |\Psi_T({\bf R})|$, 
the initial $(\tau=0)$ configuration 
${\bf R}_{0}$ for each path is obtained, as in
VMC, by sampling $\Psi_{T}^{2}({\bf R})$ using the Metropolis method.  The
subsequent configurations ${\bf R}_{i}$, at $\tau = i\triangle\tau$, are 
obtained sequentially from
${\bf R}_{i-1}$, by iterating with the importance-sampled Green's function
$G_I$,
\begin{eqnarray}
I({\bf R}_i) \Psi({\bf R}_{i}) =  \int G_{I}({\bf R}_{i},{\bf 
R}_{i-1}) I({\bf R}_{i-1}) \Psi({\bf R}_{i-1}) d{\bf R}_{i-1}.
\label{eq:import_psi}
\end{eqnarray}
This equation is the importance-sampling generalization of the iterative
form of Eq.(\ref{eq:gfmcpsi}),
\begin{eqnarray}
\Psi({\bf R}_{i}) = \int G({\bf R}_{i},{\bf R}_{i-1})\Psi({\bf R}_{i-1})d{\bf 
R}_{i-1}.
\end{eqnarray}

Eq.(\ref{eq:import_psi}) describes the evolution of the density $I({\bf 
R}_{i}) |\Psi({\bf R}_{i})| $ with $\tau = i \Delta \tau$, hence
the configurations {${\bf R}_i$} are distributed with this density.  
The propagation is entirely in
terms of distinguishable `Boltzmann' particles;
the Fermi or Bose character of the
system is retained only at the two ends of the walk through
the statistics of the initial and 
final trial wave functions. 

Up to this point, we have assumed that we can sample points along the path
directly from $G_I$, but typically this is not possible.  One
must sample from an approximate 
$\tilde{G}_I ({\bf R}_i, {\bf R}_{i-1})$
and then use the weighting
and branching techniques discussed below to create paths with probability 
proportional to the product of the $G_I({\bf R}_i, {\bf R}_{i-1})$.
If points are sampled from an approximate $\tilde{G}_I$, it is convenient to
define a weight
\begin{equation}
\tilde{W}  ({\bf R}_i, {\bf R}_{i-1}) =
\frac{G_I ({\bf R}_i, {\bf R}_{i-1})  }
{\tilde{G}_I ({\bf R}_i, {\bf R}_{i-1}) }
\end{equation}
as the ratio of the full $G_I$ to the approximate $\tilde{G}_I$.
Simply choosing paths with $ P({\bf P})  =  \prod_{i=1,n} 
\tilde{G}_I({\bf R}_i, {\bf R}_{i-1}) I({\bf R}_0)
| \Psi_T ({\bf R_0})|$ would modify
expressions for the the numerator and denominator,
Eqs.(\ref{eq:numerator}-\ref{eq:denominator}),
by multiplying the contribution of each path
by the product of the $\tilde{W}$:
\begin{equation}
N_{{\bf P}}  =  
W({\bf P})
\frac{\Psi_{T}({\bf R}_{n})\hat{O}}
{I({\bf R}_{n})}
\frac{\Psi_{T}({\bf R}_{0})}
{|\Psi_{T}({\bf R}_{0})|}~, 
\label{eq:wnumerator}
\end{equation}
\begin{eqnarray}
D_{{\bf P}} = 
W({\bf P})
\frac{\Psi_{T}({\bf R}_{n})}{I({\bf R}_{n})}
\frac{\Psi_{T}({\bf R}_{0})}{|\Psi_{T}({\bf R}_{0})|} ~,
\label{eq:wdenominator}
\end{eqnarray}
where
\begin{equation}
W({\bf P}) =
 \prod_{i=1,n} \tilde{W} ({\bf R}_i, {\bf R}_{i-1}) ~.
\end{equation}
As a trivial example, one could sample the free-particle
propagator ($\tilde{G}_I = G_0 $), and the weights $\tilde{W}$ would be
the ratio of final to initial importance functions times the ratio of
interacting to free-particle propagators.  Such a scheme, however, is woefully
inefficient.  As the path length increases, so do the fluctuations in the
W({\bf P}), and the branching techniques discussed below must 
eventually
be used to control them.

For an efficient and unbiased calculation, it can be very important
to choose a $\tilde{G}$ to minimize fluctuations in the weights $\tilde{W}$
introduced at each step.
For scalar problems,
one typically samples a shifted Gaussian, where
the shift is related to the logarithmic derivative of the trial wave
function.  This can be used to perform importance sampling accurate
to second order in $\triangle\tau$, (for a review, see Ref.
\cite{HLR94})    and hence essentially
set $\tilde{G}_I = G_I$.

In the nuclear case, though, the wave function consists of many spin-isospin
amplitudes, and a more complex sampling scheme is required.
For illustrative purposes,
we describe the scalar equivalent of our sampling method,
although for scalar interactions it is not as
efficient as sampling a shifted Gaussian.
The free propagator, 
$G_{0}({\bf R}^{\prime},{\bf R}_{i-1})$, can
be easily sampled.  A number of points, ${\bf R}^{\prime}_{j}$, 
$j=1,n_{samp}$ are obtained by sampling 
$G_{0}({\bf R}^{\prime}_{j},{\bf R}_{i-1})$.
These points should be chosen in a correlated manner to reduce fluctuations.
Anticipating requirements for the non-scalar case, we define an approximate
scalar importance-sampled Green's function $G_I^S ({\bf R}_i, {\bf R}_{i-1})$.
The primary requirements are that $G_I^S$ is fast to compute, 
that it is positive, and that it
approximates $G_I$; for the scalar case one could simply choose $G_I^S = G_I$.

For each of the $n_{samp}$ points, we calculate
$G_{I}^S({\bf R}^{\prime},{\bf R}_{i-1})$. The
${\bf R}_{i}$ is picked from the set ${\bf R}^{\prime}_{j}$ with
probability proportional to 
$G_{I}^S({\bf R}^{\prime}_{j},{\bf R}_{i-1}) /
G_{0}({\bf R}^{\prime}_{j},{\bf R}_{i-1})$.
This procedure implicitly defines a $\tilde{G}_I$,
and requires a weight
\begin{eqnarray}
\tilde{W} ({\bf R}_i, {\bf R}_{i-1})  = 
\left[\frac{1}{n_{samp}}\sum_{j=1,n_{samp}} 
\frac{ G_{I}^S({\bf R}^{\prime}_{j},{\bf R}_{i-1})}
{G_{0}({\bf R}^{\prime}_{j},{\bf R}_{i-1})} \right]
\frac{G_I({\bf R}_i, {\bf R}_{i-1})}
{G_I^S({\bf R}_i, {\bf R}_{i-1})}.
\label{eq:normweight}
\end{eqnarray}

Only the variance, and therefore the statistical sampling error in the 
calculation depends on $n_{samp}$.  When $n_{samp}=1$, 
${\bf R}_{i}={\bf R}_{1}^{\prime}$ and the vector 
${\bf R}_{i}-{\bf R}_{i-1}$ can be in any direction since $G_{0}$
depends only upon $({\bf R}_{1}^{\prime}-{\bf R}_{i-1})^{2}$.  In this
case, the weights $\tilde{W} = G_I / G_0$  
can differ significantly from unity and
add to the variance.  Indeed the growth estimate of the energy,
obtained from the difference between unity and the ratio of new to old weights,
will have an infinite variance in the limit $\Delta \tau \rightarrow 0 $.
In the present calculations, we consider only two points
${\bf R}_{1}^{\prime}$ and 
${\bf R}_{2}^{\prime}=2{\bf R}_{i-1}-{\bf R}_{1}^{\prime}$
symmetric about ${\bf R}_{i-1}$.  The leading gradient contribution along
${\bf R}_{1}^{\prime}-{\bf R}_{i-1}$, in the expansion:
\begin{eqnarray}
\frac{G_{I}({\bf R}^{\prime}_{j},{\bf R}_{i-1})}{G_{0}({\bf R}^{\prime}_{j},{\bf 
R}_{i-1})} = 1 + ({\bf R}^{\prime}-{\bf R}_{i-1})\cdot\nabla_{{\bf 
R}^{\prime}}\left(\frac{G_{I}({\bf R}^{\prime}_{j},{\bf R}_{i-1})}{G_{0}({\bf 
R}^{\prime}_{j},{\bf R}_{i-1})}\right) + \cdots
\end{eqnarray}
is thus cancelled up to order $\Delta \tau$ and the variations in 
$W ({\bf P})$
are reduced significantly.

Nevertheless, the weights of different paths used in computing expectation 
values, Eqs.(\ref{eq:wnumerator}-\ref{eq:wdenominator}) will eventually diverge.
This divergence yields an increasing statistical error, as
the contribution of only a few paths will dominate the others.
Consequently, branching techniques are required to control the fluctuations 
in the relative contributions of different paths. 
In branching, the configurations are redistributed every few time steps 
by keeping $n_{{\bf R}}^{i}$ unit weight copies of each
configuration where
\begin{eqnarray}
n_{{\bf R}}^{i} = int(W ({\bf P}) + \zeta_{{\bf R}})~,
\label{eq:branch}
\end{eqnarray}
$\zeta_{{\bf R}}$ is a random number between 0 and 1,
and $int$ denotes the (truncated) integer part.  
The $W ({\bf P})$ of the resulting configurations are then set to one
in order to account for the branching process.
This branching technique, in effect, forces the paths to be sampled
from the product of $G_I$ rather than $\tilde{G}_I$.

On average, the expectation value of any path is reproduced correctly
using this technique.
However, the computation is much more efficient as
configurations with small weights are more likely be discarded while
configurations with large weights are replicated.  In this way, statistical
noise is reduced by keeping an adequate population of contributing
configurations.

The algorithm used for nuclear GFMC, 
in which there is a strong spin-isospin dependence to the 
interaction, is a generalization of the procedure described above.
Here wave functions must be regarded as vectors in spin-isospin space and the 
$G({\bf R},{\bf R}^{\prime})$ as a matrix, however, the
relative probability of the paths $P({\bf P})$ must remain a scalar.  
Following Refs.~\cite{C87,C88,C90},
we define an importance-sampled Green's function $G_I$
as well as an approximate $\tilde{G}_I$. Just as in the scalar
case, the approximate $\tilde{G}_I$ is used for sampling points in the
path, while the ratios $G_I / \tilde{G}_I$ define weights
$\tilde{W}$ which are used in branching.

In order to introduce importance sampling, we first 
define a scalar function $I$ of the trial and GFMC, Eq.(\ref{eq:gfmcpsi}),
wave functions:
\begin{equation}
I \left[ \Psi_T({\bf R}_i), \Psi_i ({\bf P}_i ) \right] =
|\sum_{\alpha} \Psi_{T,\alpha}^{\dagger}({\bf R}_{i})
{\Psi}_{i,\alpha}({\bf P}_{i})| + 
\epsilon \sum_{\alpha} |\Psi_{T,\alpha}^{\dagger}({\bf 
R}_{i}){\Psi}_{i,\alpha}({\bf P}_{i})|~,
\label{eq:importancedef}
\end{equation}
where $\Psi_T$ is the trial wave function,
$\Psi_i$ is the $i$'th iteration of the GFMC wave function,
which depends implicitly upon the path ${\bf P}_i$, and $\alpha$ denotes
the spin-isospin components.

This definition of the importance function
differs slightly from the scalar case, which
only involved the trial function $\Psi_T$.
The first term simply measures the magnitude of the overlap
of the wave functions, while the second, with a small
coefficient $\epsilon$ ($\approx$ 0.01)  ensures a positive
definite importance function to allow diffusion across nodal surfaces.  
In this definition of $I$ as well as the remaining discussion,
we suppress the sampling of the
pair orders in the wave function and the propagator.

The importance-sampled Green's function $G_I$ can then be
defined as the ratio of the importance functions after one
iteration of the full propagator $G ({\bf R}_i, {\bf R}_{i-1})$,
\begin{equation}
G_I ( {\bf R}_i, {\bf R}_{i-1} ) =
\frac{ I \left[ \Psi_T ({\bf R}_i), \Psi_i ({\bf R}_i)\right]}
{ I\left[ \Psi_T ({\bf R}_{i-1}), \Psi_{i-1} ({\bf R}_{i-1})\right]}~,
\end{equation}
where
\begin{equation}
\Psi_i ({\bf R}_i) = G ({\bf R}_i, {\bf R}_{i-1}) \Psi_{i-1} ({\bf R}_{i-1})~.
\end{equation}
For scalar interactions, this definition of $G_I$ is equivalent to 
Eq.(\ref{eq:importancegf}). Here, the importance function is defined
from all the amplitudes of the trial and GFMC wave functions, and
the effects of the propagator are included in the importance function.

 To perform a calculation, the initial configurations are sampled
from $I({\bf P}_0)$, which is defined by inserting $\Psi_{i=0} = \Psi_T$ 
in Eq.(\ref{eq:importancedef}) above.
For speed, the VMC calculations use the simple importance
function $W_{pq}({\bf R})$ defined in Eq.\ref{eq:vmc:weight}.
Hence we introduce the ratio of the
two importance functions as an initial weight and perform a branching
step immediately.  This procedure results in a population
drawn from $I({\bf P}_0)$.

Sampling steps from $G_I$ gives us
\begin{eqnarray}
P({\bf P}) & = & \prod_{i=1,n} 
\frac{I \left[ \Psi_T ({\bf R_i}), \Psi_i ({\bf R}_i) \right]}
{I \left[ \Psi_T ({\bf R_{i-1}}), \Psi_{i-1} ({\bf R}_{i-1}) \right]}
I \left[ \Psi_T ({\bf R_{0}}), \Psi_i ({\bf R}_{0}) \right] \\
& = & I \left[ \Psi_T ({\bf R_n}), \Psi_n ({\bf R}_n) \right]~,
\end{eqnarray}
and hence estimates of observables as the ratio $N/D$ where
\begin{eqnarray}
N_P & = & \frac{  \Psi_T^{\dagger} ({\bf R}_n)  O \  \prod_{i=1,n} 
G({\bf R}_i, {\bf R}_{i-1})  \Psi_0 ({\bf R}_0) }
{ I \left[ \Psi_T ({\bf R_n}), \Psi_n ({\bf R}_n) \right]}~, \nonumber \\
D_P & = & \frac{ \Psi_T^{\dagger} ({\bf R}_n)  \prod_{i=1,n} 
G({\bf R}_i, {\bf R}_{i-1})  \Psi_0 ({\bf R}_0) }
{ I \left[ \Psi_T ({\bf R_n}), \Psi_n ({\bf R}_n) \right]} ~.
\label{eq:nucobserv}
\end{eqnarray}
For scalar interactions, setting $\epsilon = 0$ in the definition of $I$, 
Eq.(\ref{eq:importancedef}), we recover 
Eqs.~(\ref{eq:numerator}-\ref{eq:denominator}).

Again, though, we cannot sample from $G_I$ directly.  We must sample
from a $\tilde{G}_I$ and introduce weights and branching to correctly
get paths sampled from the products of $G_I$. The procedure is
exactly as described previously, although
here
it is important to
introduce an approximate $G_I^S$.
The $G_I^S$ is a spin-independent function, and hence is
much easier to compute
than $G_I$, which involves all the spin-isospin states of the system.
The present algorithm
requires us to compute only a single full propagator per iteration,
and the full trial wave function only after several iterations.

The scalar importance function
${G}_I^S ({\bf R}',{\bf R}_{i-1})$
is again used to implicitly define 
$\tilde{G}_I$, and construct weights
$\tilde{W}_I ({\bf R}_i,{\bf R}_{i-1})$.
It contains scalar approximations to the dominant physics present in the
propagator and the trial wave function,
\begin{equation}
G_{I}^S ({\bf R},{\bf R}^{\prime}) =
|\Psi_J ({\bf R})| G^S ({\bf R}, {\bf R'}) \frac{1}{|\Psi_J ({\bf R'})|}~,
\end{equation}
where $G^S ({\bf R},{\bf R}^{\prime})$ is obtained from
an approximate spin-isospin independent interaction:
\begin{eqnarray}
v_{S}(r_{ij}) = \frac{1}{2}[v_{c}(^{1}S_{0},r_{ij}) + v_{c}(^{3}S_{1},r_{ij})]
\end{eqnarray}
and the Feynman approximation
\begin{eqnarray}
G^S({\bf R},{\bf 
R}^{\prime})=\exp(-\sum_{i<j}v_{S}(r_{ij})\triangle\tau/2)G_{0}({\bf R},{\bf 
R}^{\prime})\exp(-\sum_{i<j}v_{S}(r^{\prime}_{ij})\triangle\tau/2)~.
\end{eqnarray}
This propagator uses the average of the central potentials in the 
important S-waves, and,  
like the true $G({\bf R},{\bf 
R}^{\prime})$,
is small at small $r_{ij}$ , preventing the configurations from
having small inter-particle distances inside the repulsive core range.
Similarly, the approximate importance sampling in $G_I^S$ is governed
by the function $|\Psi_J|$, Eq.(\ref{eq:jastrow}), which can be used as a 
simple approximation to $|\Psi_{T}|$.

With these definitions, a step in the propagation is the same as in the
scalar case.
It begins with sampling $n_{samp}$ 
correlated points ${\bf R}'_j$
from the free-particle propagator.  Then
${G}_I^S$ is evaluated for each possible step, and we choose
${\bf R}_i$ from them.  The weight
$\tilde{W}$
is then computed as the ratio of importance functions divided by the sampling
probability $\tilde{G}$ as in Eq.(\ref{eq:normweight}).

Since we typically do not compute observables after each step,
and fluctuations in the weights are not significant after a single step,
it is not necessary to compute the importance function $I$ 
(and hence the trial wave function) at every step.
We perform branching after every second step, with the weights computed
from the product of intermediate $\tilde{W}$.  This product involves
only the $G_I^S$ and overlaps of the wave functions at the final
and initial steps. Hence, for intermediate steps we must
compute the full $G$ acting on the GFMC wave function, but not necessarily the
full trial wave function. Again, after branching, the weight of each path is
set to 
$W({\bf P}) = 1$.

At this point we can reconstruct the estimates of any observables.
After branching,
the $N_{{\bf P}}$ and $D_{{\bf P}}$ obtained with this $P({\bf P})$ are given 
by Eq.(\ref{eq:nucobserv}).
Ignoring $\epsilon$, the variance of $D_{{\bf P}}$ is mostly due to the
Fermion sign problem, while that of $N_{{\bf P}}$ is tolerable, particularly
when $\hat{O}={H}$ and $\Psi_{T}$ is close to the desired eigenfunction
of ${H}$.  For an exact $\Psi_T$, we regain the exact ground state
energy with zero variance.

\section{COMPUTATIONAL METHODS}

	The complicated nature of the nuclear interaction and the
computational complexity of the calculations presented here require
high performance computing.  In the past, vector supercomputers
were used for the first $^{4}$He and $^{5}$He GFMC calculations.
As we have stated before, the size of the wave function vector grows 
exponentially with the number of nucleons and the number of
matrix operations grows with the number of pairs.  In
making the step from four- to six-body calculations at least
an order of magnitude increase in computational performance 
was required.  The clear means of achieving this performance
goal was parallel computation.   

	A frequent method of achieving performance gains in 
Monte Carlo calculations is to distribute the 
configurations over several processors and let each processor
carry out its own independent Monte Carlo calculation.  Such an 
``embarrassingly parallel'' implementation is sufficient for simple 
calculations in which each processor can handle a calculation
with a minimum acceptable number of configurations in a
reasonable amount of time.  For the seven-body systems considered
here this is not the case.

	The heart of the VMC calculation is
the Metropolis algorithm which is an inherently serial algorithm.
Since the bulk of the work in our variational calculations lies
in the energy expectation value, the straightforward division
of labor is to have one master processor perform the Metropolis walk
while several slave processors calculate the energy and other
expectation values for the configurations that the master generates. 
The number of slave processors that can be efficiently used is the ratio
of the CPU time needed for expectation values to that needed to walk
from one configuration to the next.  We find that typically 50
processors can be used efficiently in a $^{7}$Li VMC calculation.

Implementing the GFMC algorithm on a parallel architecture provides 
some special challenges.  There is no clear division of labor as
in the VMC calculation and the number of configurations can change
throughout the calculation.  An embarrassingly parallel implementation
could work for $^{6}$Li, since one processor on current machines
is capable of handling enough $^{6}$Li configurations (several thousands)
for an independent GFMC calculation.  For $^{7}$Li, on the other hand,
only hundreds of configurations could be propagated on each processor
to achieve an acceptable turnaround time for an independent calculation.
With such a small configuration set on each node, the population 
fluctuations on each processor would leave some processors with few
configurations and others with too many.  To avoid such an inefficient
use of resources, periodic load balancing between processors is 
required.
	
In our implementation, the initial configurations are generated and written
to disk in a random walk that uses only two processors.  Typically 50,000
configurations are generated.  These are then used in one or more
subsequent GFMC calculations.  At the start of the GFMC calculation, the
master processor reads the configurations and distributes them to
the slave processors.  It is then responsible for collecting and averaging
energy expectation values, and determining load-balancing distributions.
Each of the slave processors is responsible for a block of configurations.
The slaves perform propagation and branching for 
this block of configurations.  At selected values of
$\tau$ (typically every 20 steps) they save the configurations in local lists
for subsequent energy calculations.
During load balancing (which typically is done every 10 steps)
each slave reports its current load to the master, which then
instructs each over-loaded slave to send its excess 
configurations to under-loaded slaves.  In this way, all slaves 
have, within tolerances, the same number of configurations.  
The master must receive load information from all the slaves before it
can determine the redistribution.  
In order that slaves not remain idle while waiting
for the redistribution information, they compute energies for the
configurations stored in their local lists.
When a slave 
completes a block of energy calculations for a particular time step, they 
are sent back to the master. After all of the energy results 
for a time step are received by the master, they are averaged.
This program structure scales well with the number of processors since 
there are no major communication bottlenecks in the course of the 
calculation and load balancing keeps the slaves somewhat synchronized.
Calculations with up to 50 processors on the Argonne IBM SP show no
degradation in efficiency; typically the slaves are idle less than 5\% 
of the time and most of this idle time occurs at the end of the calculation.

The current version of our GFMC program is written using Fortran 90 and
makes use of the MPI message-passing library.  
On IBM SP1 processors, we achieve 40\% (33\%) of the theoretical speed 
in $^{6}$Li ($^{7}$Li) calculations.  On IBM SP2 wide nodes we get ~45\%
for both cases.  The better efficiency can be ascribed to the larger
cache on the wide nodes.
Table~\ref{table:perform} shows the performance of our GFMC program on the 
Argonne IBM SP (using SP1 nodes) and the Cornell IBM SP2 using wide nodes.  
The table gives the CPU times for a single propagation step and a single
energy evaluation and the total time required for
50,000 configurations propagated for 120 time steps 
($\triangle\tau$=0.0005 MeV$^{-1}$, $\tau_{final}$=0.06 MeV$^{-1}$) 
with load balancing every 10 time steps and energy evaluations every
20 steps; this time includes slave idle time and average effects of
configuration number growth.  It is based on results using 20--40 slaves.  
However, this total time does not include the master time.
As one can see, the time grows by roughly a factor of 10 
from $^{4}$He to $^{6}$Li to $^{7}$Li.  
The total computational effort approximately scales as the product of the wave 
function size $N(A,T)=2^A \times I(A,T)$ [Eq.~(\ref{eq:numiso})]
and the number of pairs $P=\case{1}{2}A(A-1)$.
This rapid increase in computation time will be a serious obstacle to extending
these calculations beyond eight-body nuclei on presently available computers.

Our earlier 6-body calculations \cite{PPCW95} were about 30 times slower
than those we now make.
The increased speed is approximately attributable to 
1) using the exact two-body propagator (factor of 3), 
2) dropping the $\epsilon_{C}V_{ijk}^{C}$ term in the $\Psi_{T}$ (factor 2),
3) improvements in the calculation of the wave functions (factor 3), and
4) reduction of processor idle time (factor 1.4).

\section{ACCURACY OF GFMC}

In this section we consider several aspects of our calculations that
could introduce systematic errors in the GFMC results, and attempt to
place limits on these errors.  There are two major sources of error: due
to the fermion sign problem, the GFMC propagation cannot be extended to
arbitrary large imaginary time and thus admixtures of low-energy excitations 
in the trial wave function will not be fully removed, and the GFMC propagation
must be done with a different Hamiltonian from the desired one.  We also
investigate the effects of the time step size.

Figure~\ref{fig:sigma_of_tau} shows the statistical errors as a function
of imaginary time for calculations of $\langle H \rangle$ for various
nuclei using 50,000 initial configurations. The 
errors grow dramatically with increasing $A$ due to the increasingly
poorer quality of the $\Psi_{T}$.  In these calculations, $\langle
\Psi_{T}|H|\Psi (\tau)\rangle$ is evaluated by having $H$ act entirely to
the left, therefore if $\Psi_{T}$ is an eigenfunction of $H$, the
sampling errors will be nil as per Eq.~(\ref{eq:expectation}).  
For the $p$-shell nuclei, the errors increase exponentially with
$\tau$.  This is the well-known fermion sign problem; the $s$-shell nuclei
suffer much less from this problem~\cite{CFP88}.  This
exponential error growth places an effective limit of
$\tau_{final}$=0.06 MeV$^{-1}$ on our calculations for $A=7$ nuclei,
which means that admixtures of states of excitation energy less than 
$\sim$12 MeV in $\Psi_T$, will be damped by less than 50\%.

This led us in Ref.~\cite{PPCW95} to attempt to extrapolate the
computed $E(\tau)$ to $\tau = \infty$ by fitting them with
\begin{equation}
E(\tau) = E_0 + \frac{ \sum_{i} \alpha_i^2 E_i^\star exp(-E_i^\star \tau) }
{1 + \sum_{i} \alpha_i^2 exp(-E_i^\star \tau) } \ ,
\end{equation}
where $E_0$ is the extrapolated energy, and the strengths of
contaminating states in $\Psi _T$ are approximated with a few $\alpha_i^2$
at excitation energies $E_i^\star$.
Figure~\ref{fig:he4_e_tau} shows such fits made for $^4$He.  Because $^4$He
calculations are so inexpensive, we were able to make precise
calculations of $E(\tau)$
for many values of $\tau$ and thus determine the short-time
behavior of $E(\tau)$ using 200,000 to 740,000 configurations.  
The solid curve is a fit with $E_0 = -28.335$ MeV, excitation
energies $E_i^\star = 20.2$, 341, and 1477 MeV, and corresponding 
$\alpha_i^2 = 0.0062$, 0.0018, and 0.00046.  
The lowest $0^+$ excitation of $^4$He is
at 20.2 MeV and this energy was not varied in the fit.  The $\chi^2$
of the fit is 19 for 31 $E(\tau)$ (25 degrees of freedom), 
so the $E(\tau)$ are not statistically independent.  
We have not attempted to estimate
the correlations between the energies at different times.
The $\chi^2$ increases by 1 when $E_0$ is changed by +0.02 or $-0.03$.
The dashed curve shows a fit without the 20.2 MeV excitation; it gives
$\chi^2 =$ 23 and $E_0 = -28.28$ MeV.  
For most of the other GFMC calculations reported in this paper, we did not
compute $E(\tau)$ at the many $\tau < 0.1$~MeV$^{-1}$ used in these fits.  Therefore
we made several fits to the  $E(\tau)$ for  $\tau \geq 0.1$.
A fit using  $E_1^\star = 20.2$~MeV and 
one adjustable  $E_i^\star$ gives $E_0 = -28.33(3)$ with $\chi^2 = 14.6$
(11 degrees of freedom), while a fit with just one $E_i^\star$ results in
$E_0 = -28.33^{+.04}_{-.12}$, $E_1^\star = 30$, and $\chi^2 = 16.0$.
Finally, the dash-dot line and
dotted lines shows the average of the 
$E(\tau)$ for $0.04 \leq \tau \leq 0.1: -28.300(15)$.  
It appears that in this most favorable case, with high statistics,
high first excited state, and large maximum $\tau$,
we can see that including the first excited state improves
the extrapolation marginally.   
However, the extrapolated $E_0$ is not
significantly lower than a simple average of the $E(\tau)$ for 
$0.04 \leq \tau \leq 0.1$.

Figure~\ref{fig:li6_e_tau} shows the $E(\tau)$ and fits made for the
ground state of $^6$Li.  The values for $\tau > 0.06$ MeV$^{-1}$ were
computed with 200,000 initial configurations, those for $\tau = 0, 0.01,
..., 0.06$~MeV$^{-1}$ have 280,000 configurations, while those for the
other small $\tau$ have only 50,000 configurations.  
The energy at very small $\tau$ is influenced by admixtures of very
high energy states in $\Psi _T$.  These have little effect on the
$E(\tau > 0.1$~MeV$^{-1}$, therefore we make fits to $E(\tau)$ only for
$\tau > 0.01$.  The dashed curve is a fit to the $E(\tau)$ for $0.01
\leq \tau \leq 0.06$, which is the range that is available for the other 
$p$-shell nuclear states in this paper.  The extrapolated energy is $E_0 =
-31.56^{+0.24}_{-0.50}$ MeV, where the indicated errors correspond to
$\chi^2$ increasing by 1.  This fit was made using a single excitation
energy, $E_1^\star = 36$ MeV.  The first 1$^+$ excited state of $^6$Li
is at 5.65 MeV.  A single-energy fit constrained to this energy gives
large $\chi^2$.  Two-energy fits with one energy constrained to 5.65 MeV
have a very flat $\chi^2(E_0)$ from which useful values of $E_0$ can 
not be extracted.  
The solid curve shows a single-energy fit made to the
$E(\tau)$ up to 0.1~MeV$^{-1}$ available for this state
; it gives $E_0 = -31.38^{+0.12}_{-0.18}$.  We see
that including data up to 0.1~MeV$^{-1}$ reduces the error in $E_0$ by
about a factor of two.
Finally the solid line with dotted error bars is the
average of the $\tau =$ 0.04, 0.05, and 0.06 values, denoted 
by $E_{av}$. Its value, -31.25(11) MeV, is formally an upper bound
for $E_0$ and is above the extrapolated
$E_0$ by only one standard deviation.

Because of the difficulties in making useful extrapolations in $\tau$, it is 
important to understand contaminations in $\Psi_T$,
particularly from low-excitation-energy states which will not be fully
filtered out by $\tau = 0.06$ MeV$^{-1}$. 
We have made several calculations of the ground-state of $^6$Li
to study the effects of changes in $\Psi_T$ on the GFMC
$E(\tau)$.  Figure~\ref{fig:li6_trunc} shows the
effects of removing some of the noncentral correlations in $\Psi_T$;
the solid circles are from a calculation with
the full $\Psi_T$ and are the same as in Fig.~\ref{fig:li6_e_tau}.  The
open squares were computed by using the simpler $\Psi_P$ of Eq.(\ref{eq:psip}).
This makes the energy at $\tau = 0$ worse by
$\sim 1.7$ MeV.  However by $\tau = 0.01$, the GFMC has fully corrected for
this defect and thereafter the differences are just statistical
fluctuations.  Hence removing $\tilde{U}^{TNI}_{ijk}$ from $\Psi_T$
enhances the admixtures of excitations $> 250$ MeV.  
Calculations without the $\tilde{U}^{TNI}_{ijk}$ would be about 20\%
faster than full calculations, but the poorer quality of $\Psi_T$
without $\tilde{U}^{TNI}_{ijk}$ increases the statistical errors at
large $\tau$ by about 20\%, and thus requires 40\% more configurations
to get the same error.
Therefore it is not economical to drop the $\tilde{U}^{TNI}_{ijk}$ 
from $\Psi_T$.

The open diamonds in Fig.~\ref{fig:li6_trunc} come from a much more
drastic approximation of $\Psi_T$.  Here both the $\tilde{U}^{TNI}_{ijk}$
and tensor components $U_{ij}$ have been omitted, resulting in a
four-operator wave function.  In such a wave function, the
dominant tensor components of the two-body potential have zero
expectation value and the energy at $\tau = 0$ is $+41$ MeV.
It is completely corrected by $\tau = 0.03$; again the rate of
correction indicates excitation energies $\sim 250$ MeV.  The
statistical errors from such a bad $\Psi_T$ are much larger.

These two tests indicate that defects in the non-central parts of the
correlation, which have been the subject of much optimization in VMC
studies, are easily corrected by the GFMC.
Deficiencies in the one-body part of $\Psi_T$ present more of a problem.
As is discussed in Sec. III, the $\Psi$ for $^6$He has two symmetry
components: [2] and [11]; the optimal amplitudes for these (see
Table~\ref{table:beta6}) are 0.967 and $-0.253$, respectively.  The
solid circles in Fig.~\ref{fig:he6_phis} show GFMC energies from a
$\Psi_T$ using these components.  The open squares show results computed 
using a $\Psi_T$ with just the [11] component; the $E(\tau=0)$ 
obtained with such a wave function is
4.5 MeV higher than that obtained with the best $\Psi_T$; this
corresponds to the 5 MeV excitation energy of the dominantly [11] state.
However, because this 
error is entirely due to a low energy excitation, the GFMC makes very little
improvement by $\tau = 0.05$ MeV$^{-1}$.  A less radical case is shown
by the open diamonds which correspond to a $\Psi_T$ with amplitudes of
$+\case{1}{\sqrt{2}},~-\case{1}{\sqrt{2}}$ for the two states.  The $E(\tau)$
starts out 1.6 MeV above that of the best $\Psi_T$; the GFMC  reduces this
to $\sim 1.3$ MeV at
$\tau = 0.06$ MeV$^{-1}$.  The solid curve is a single-energy fit to
these results; the fitted excitation energy is $36$ MeV and the
extrapolated $E_0 = -26.7$ is well above the $E_{av}=-27.64(14)$
from the best $\Psi_T$.  Fits with two
excitations, one constrained to 5.0 MeV, give an essentially flat
$\chi^2$, and are not useful to extract the $E_0$.

A similar situation arose in our first GFMC calculation for the ground
state of $^7$He.  This was made with just the dominant $^2$P[21]
component.  A subsequent calculation using the three components listed
in Table~\ref{table:beta7} lowered all the $E(\tau)$ by $\sim 0.8$~MeV;
the first $\case{3}{2}^-$ excited state is at
only 2~MeV so we would have to propagate 10 times further for the GFMC
to substantially correct this error.

It is important that the
$\Psi_T$ have the correct admixtures of different symmetry, and other
low-lying states.  Otherwise the GFMC results will only be upper bounds
to the exact eigenenergies.  For this reason the $\Psi_T$ used in this
work are obtained by diagonalizing the Hamiltonian in the small bases
of low-energy shell-model states as discussed in Sec. IV. 
Such diagonalizations can be used to ascertain if
an improved $\Psi^{\prime}_T$ will influence the GFMC results
obtained with $\Psi_T$.  The Hamiltonian should be diagonalized
between $\Psi^{\prime}_T$ and $\Psi_T$, taking into account their
nonorthogonality. If the difference in the eigenvalues is large compared
to $1/\tau_{final}$ the results will not be influenced;
if it is small, the superposition corresponding to the lowest
eigenvalue must be used.  
As examples of this, we made such diagonalizations for the cases studied
above, in which  $\Psi_P$ or just a four-operator $\Psi_T$ were used.
These gave excitation energies of  $\sim 700$~MeV and  $\sim 300$~MeV,
respectively, which are in good agreement with the observed $E(\tau)$.
Our best $\Psi_V$ contains $U_{ij}^{LS}$ and
$U_{ijk}$ correlations omitted from the $\Psi_T$ due to computational
costs.  Diagonalizing the Hamiltonian within $\Psi_V$ and $\Psi_T$ shows
that these correlations admix states with excitation energies of $\sim 1000$
and 300 MeV, respectively.  Thus they can be safely left out of the
$\Psi_T$.

As is discussed in Sec.~III, the VMC calculations for $p$-shell nuclei do
not have a local variational minimum for reasonable rms radii.
Therefore the variational searches were constrained to have radii close
to the experimental values, if such values are known.  To study the
sensitivity of the GFMC results to this assumption, we have made a
number of GFMC calculations of the $^6$Li ground state using $\Psi_T$
that have different rms radii.  These $\Psi_T$ were made by changing the
depth ($V_p$) and radius ($R_p$) of the Woods-Saxon well used to make
the $p$-wave orbitals (see Eq.~(\ref{eq:spwell})); thus the $^4$He core
was not directly modified.  Figure~\ref{fig:6li_rmsr} shows the
evolution of the rms radius with $\tau$ for these calculations; the
solid circles correspond to the $\Psi_T$ used in the rest of this paper.
We see that the GFMC basically makes no change to the radii, even though
they span a range of almost 30\% (there may be some indication that the
smallest radii are increased at small $\tau$).  This is probably because
completely separating the deuteron from the $^4$He core
corresponds to only a 1.5 MeV excitation.
Figure~\ref{fig:6li_e_vs_rmsr} shows the GFMC energies from these
calculations as a function of the GFMC radii (both are averages of the
$0.04 \leq \tau \leq 0.06$~MeV$^{-1}$ values), and the corresponding
$\Psi_T$ expectation values.  The variational energies obtained with
$\Psi_T$ decrease
monotonically with increasing rms radius, but the GFMC energies show a
weak minimum; the very large radii yield higher GFMC energies and
thus can be variationally ruled out.  The curve is a parabolic fit to
the five GFMC energies with smallest rms radii; the minimum is at 2.44 fm.  
However the curve is very flat and the uncertainty in the location of the
minimum is at least 0.1 fm.
Thus even when 50,000 to 280,000 initial configurations are used for 
each point, it is difficult to extract the $^6$Li rms radius.
Given this, it is not possible at present to reliably study the radii of $A=7$
nuclei.
 
The GFMC propagator used in these calculations has two possible
sources of error.  The first is that the time step, $\triangle\tau =
0.0005$ MeV$^{-1}$, might be too large.  We have checked this by two
calculations.  For $^3$H we made calculations to $\tau = 0.06$
MeV$^{-1}$ using both $\triangle\tau = 0.0005$ and 0.00025 MeV$^{-1}$;
these were different by 0.016(14) MeV.  For $^4$He we made calculations
to $\tau = 0.01$ MeV$^{-1}$ using both $\triangle\tau = 0.0005$ and
0.0001 MeV$^{-1}$; these were different by 0.02(6) MeV.  The statistical
errors in our GFMC energies for $p$-shell nuclei are all $> 0.1$~MeV; thus
these time-step effects are negligible.

The second possible source is that we cannot use a propagator for the
full Hamiltonian, $H$, that we are interested in.  Rather we must use
the $H^{\prime}$ of Eq.~(\ref{eq:hgfmc}) and compute $\langle H-H^{\prime}
\rangle$ perturbatively.  Kamada and Gl\"{o}ckle \cite{glockle_private}
have estimated for $^3$H that evaluating $\langle v_{ij} \rangle$ in
wave functions appropriate for $v^{\prime}_{ij}$ underestimates the
binding energy by $< 20$~keV; scaling this by the total two-body
potential energy gives $< 50$~keV in $^4$He and $< 90$~keV in $^7$Li.
We constructed a $^4$He $\Psi_V$
that is optimized for $v^{\prime}_{ij}+V_{ijk}$ and used it to compute
$\langle v_{ij} \rangle$; this gives 60(15)~keV less binding than our
best $^4$He variational wave function in agreement with the above
estimates. 


Using a propagator for an $H^{\prime}$ that gives more binding than $H$
can introduce small errors in the determination of densities and
radii.  The more tightly bound eigenstate of $H^{\prime}$ is likely to
have a smaller radius.  The radii of $^4$He can be calculated more
accurately, and we have studied their sensitivity to 
various propagators.  The rms radius for
$\Psi_T$ optimized for $H$ 
is 1.482(3)~fm.  A GFMC calculation using an $H^{\prime}$ with
no $v^{\prime}_{C}(r_{ij})$ and $V^{\prime}_{ijk} = V_{ijk}$ (see
Eq.~(\ref{eq:hgfmc})) gives $\langle H - H^{\prime} \rangle = 2.40(3)$~MeV
and an rms radius of 1.418(4).  However using the
$v^{\prime}_{C}(r_{ij})$ and $1.3U_0$ in $V^{\prime}_{ijk}$ results in
$\langle H - H^{\prime} \rangle = .03(2)$~MeV and an rms radius of
1.446(3)~fm. Presumably the later value is more correct, while the former is
too small due to the overbinding.

As is discussed in Sec.~IV, the GFMC directly computes mixed estimates
$\langle O (\tau)\rangle_{Mixed}$.
Except for $H^{\prime}$ and operators that commute with it, these must 
be corrected to obtain the desired $\langle O (\tau)\rangle$; we use 
Eq.~(\ref{eq:pc_gfmc}) to achieve this.
Consequently, the expectation values of the individual energy components,
such as $K^{CI}$, $v^{\prime}_{ij}$, and $V^{\prime}_{ijk}$, which have
errors of order $|\Psi _0-\Psi _T|^2$, 
do not sum to the correct total energy.
Indeed, there must be a collective error in these individual terms equal
to the total difference between the GFMC $\langle H^{\prime} \rangle_{Mixed}$
and the VMC $\langle H^{\prime} \rangle_T$.  
This is illustrated in Table~\ref{table:pc_cor} for the case of $^6$Li,
where the difference $\langle H^{\prime} \rangle_{Mixed}-\langle H^{\prime} 
\rangle_T$ is $-4.4$ MeV, and the sum of the individual $\langle O \rangle$ is
an additional $-4.4$ MeV lower than $\langle H^{\prime} \rangle$.
In this case, the individual corrections are comparable in magnitude to the 
collective error, but small compared to the total expectation values.

Aside from our own work in Ref.~\cite{PPCW95}, there are no published
calculations of $p$-shell nuclei using realistic interactions such as
those used here.  However we can compare to previous values for
the $s$-shell nuclei.
There are accurate Faddeev and projected hyperspherical harmonics (PHH)
calculations of $^3$H for the Argonne $v_{14}$ with no $V _{ijk}$.  For
this Hamiltonian we find a GFMC energy of $-7.670(8)$ MeV which is in good
agreement with the previous results of $-7.670$ (Faddeev/R
\cite{CPFG85}), $-7.680$ (Faddeev/Q \cite{G95}), and $-7.683$ (PHH
\cite{KVR94}).  A PHH result for Argonne $v_{18}$ with Urbana IX has
recently been computed\cite{viviani_private}: -8.475. It is in good
agreement with our values of -8.455(8) obtained with
$\triangle\tau$=0.5 GeV$^{-1}$ and $-8.471(12)$ obtained with 
$\triangle\tau$=0.25 GeV$^{-1}$.

There are also several other calculations of $^4$He with Argonne
$v_{14}$ without $V _{ijk}$.  Figure \ref{fig:4he_v14} shows the GFMC
$E(\tau)$ for this case.  Because there is no $V_{ijk}$, we multiplied
the $v^{\prime}_{ij}$ in $H^{\prime}$ by 0.994 and included the
isoscalar $v^{\prime}_C$ so that $\langle
H-H^{\prime} \rangle \sim 0$.  The average, shown by the line and dotted
error range, of the last few $E(\tau)$ is $-24.227(31)$ MeV.  A
calculation using the full $v^{\prime}_{ij}$, without isoscalar
$v^{\prime}_C$, in $H^{\prime}$, gives $-24.230(31)$, even though in this
case $\langle H-H^{\prime} \rangle = 1.5$ MeV.  These results are in
excellent agreement with the older GFMC calculation of
Ref.~\cite{CS94} (up-pointing triangle) which was made
with a completely independent program that uses the short-time
propagator of Eq.~(\ref{eq:propagator1}).  They are also in excellent
agreement with the correlated hyperspherical harmonic (CHH) value of
$-24.17\pm{.05}$~MeV\cite{viviani_in_prep} shown by the open diamond.
The error bar on the CHH value represents the expected truncation error
in that calculation.  However these results are below the Coulomb-corrected
Faddeev-Yakubovsky value of $-24.01$ MeV\cite{G95}, shown by the open
circle.  The down-pointing triangle shows our best VMC upper bound for
this case; it is $\sim 2$ \% higher than the exact $E_0$.

In the following sections we will give the calculated values of
the average energy ($E_{av}$) for $\tau =$~0.04, 0.05, and 0.06 MeV$^{-1}$
for various A = 6 and 7 states.  These provide upper bounds to the
eigenenergies of the nuclear Hamiltonian $H$ used in this work.  The
studies of the accuracy of GFMC discussed above suggest that, assuming
that the low-energy excitations in the $\Psi _T$ have been successfully
removed by diagonalizing the Hamiltonian matrix in the $p$-shell states,
these $E_{av}$ are at most $\sim 0.3$ MeV above the eigenenergies
for A = 6 states. For the $^6$Li ground state the additional binding obtained by
single energy extrapolations is only $\sim 0.13(15)$~MeV.
This extrapolation was made using a factor five more samples than
we have for the other states studied, thus no useful extrapolation
estimates can be made for the other states.
We also estimate that the
perturbative treatment of $H-H^{\prime}$ increases $E_{av}$ by less than
0.1~MeV.  As will be shown in the next section, the
large $\tau$ behavior of $E(\tau)$ for all the A = 6,7 states is
very similar.  Thus we expect that the errors
estimated for $^6$Li are reasonable approximations to those for other nuclei. 
By scaling the above two errors according to $E(\tau=0)-E_{av}$ and
$\langle v_{ij} \rangle$ we estimate that the $E_{av}$ for the $A = 7$
states is no more than $\sim 0.5$ MeV above the eigenenergies.

\section{ENERGY RESULTS}

\subsection{Ground States}

The primary results of this paper are the GFMC energies, $E_{av}$, of the ten 
different $(J^{\pi};T)$ states in $A=6,7$ nuclei shown in 
Table~\ref{table:energy} and in Fig.~\ref{fig:energy_spectra}, along with
three isobaric analog states and the ground states of $A=2-4$ nuclei. 
The present results for $^6$He and $^6$Li ground states and the (3$^+$;0)
excited state in $^6$Li are all slightly lower, but within error bars, of
the $\tau$-averaged results reported in Ref.~\cite{PPCW95}.
The slight improvement may be due to the improved $\Psi_T$, while we have 
obtained much better Monte Carlo statistics than previously, both by 
access to increased computer resources and by more efficient program
implementation.
The $\tau$-extrapolated results of Ref.~\cite{PPCW95} were significantly
below the present results, but also had a large uncertainty associated with
the extrapolation; taking that uncertainty into account, the two calculations
are consistent.
In the particular case of the $^6$Li ground state, our earlier result of
$-32.4(9)$ MeV was not inconsistent with the experimental binding of 
$-31.99$ MeV.
However, from the various extrapolation tests discussed above, and the
propagation to $\tau = 0.1$ MeV$^{-1}$, we are now confident 
that the binding energy with the present Hamiltonian is not more than $-31.6$ MeV.

Table~\ref{table:energy} also gives the VMC energies from the simple 
starting trial function, $\Psi_T$, and from the more sophisticated $\Psi_V$.
In $A=3,4$ nuclei, $\Psi_V$ picks up about 60\% of the energy difference
between $\Psi_T$ and the final GFMC results.
However, for the $A=6,7$ nuclei there is a much bigger gap between VMC
calculations with $\Psi_T$ and the GFMC energies, and the $\Psi_V$ results
recover only 20--25\% of this energy difference.
Clearly there is some important aspect to $p$-shell variational wave functions
that is missing from the current {\it ansatz}.

The Argonne $v_{18}$ + Urbana IX Hamiltonian was constructed to reproduce the 
experimental binding energies of $^2$H, $^3$H, and the equilibrium
density of nuclear matter.
From Table~\ref{table:energy} and Fig.~\ref{fig:energy_spectra} we see that 
with this Hamiltonian all the $A=6$ and 7 states studied here are underbound.
The discrepancy in $^6$Li and $^7$Li states is relatively small, $<2\%$ and $<5\%$ 
respectively, and the calculated ground states are stable against breakup 
into $\alpha + d$ and $\alpha + t$.
On the other hand, the discrepancy in $^6$He and $^7$He states is larger,
$\sim 5\%$ and $\sim 13\%$, respectively, and the calculated $^6$He ground
state is unstable against $\alpha + n + n$ breakup.

A breakdown of the GFMC energies into kinetic and potential contributions
is given in Table~\ref{table:detail}.
The kinetic and potential energies grow rapidly as the number of nucleons 
increases, but for a given nucleus, they decrease slightly as the excitation 
energy increases and the nucleus gets more diffuse.
The $V_{ijk}$ contribution remains small compared to $v_{ij}$, never exceeding
5\%, but because of the large cancellation between $K$ and $v_{ij}$, it
is typically 25\% of the total binding energy.
The electromagnetic $v^{\gamma}_{ij}$ is dominated by
the Coulomb interaction between protons, $V_{C1}(pp)$, but about 17\% (8\%)
of its total contribution comes from the magnetic moment and other terms in
Eqs.(\ref{eq:vempp}-\ref{eq:vemnn}) in He (Li) isotopes.
The one-pion-exchange term of the potential dominates $v_{ij}$, providing
$\sim 70\%$ of the interaction energy, while the $V^{2\pi}_{ijk}$ is
smaller than $v^{\pi}_{ij}$ by one order of magnitude.

Figure \ref{fig:gfmc-vmc} shows the $E(\tau)-E(\tau\!\!=\!\!0)$ from the GFMC
calculations of ground states of nuclei with $3 \leq A \leq 7$.  The
GFMC correction to the VMC ($\Psi_T$) results has a strong $A$
dependence but no significant $N-Z$ dependence.  
Figure \ref{fig:normalized_gfmc-vmc} shows $[E(\tau)-E(0)]/|E_{av}-E(0)|$.
The results for the two $s$-shell nuclei have
the same dependence on $\tau$, as do those for the four $p$-shell
nuclei.  However the $p$-shell $E(\tau)$ approach their asymptotic values
less rapidly; a fit to the $^6$Li $E(\tau)$ for $\tau < 0.03$~MeV$^{-1}$
requires excitation energies of $\sim$~700 and 90~MeV instead of the
$\sim$~1500 and 350 MeV used in the fit for $^4$He shown in
Fig.~\ref{fig:he4_e_tau}.  This is another indication that there is a
qualitatively new feature necessary for $p$-shell nuclei which is
missing from our trial wave functions; this feature does not seem to
depend on the $N-Z$, $J^{\pi}$, or $T$ of the nucleus.

\subsection{Excited states}

A second result of the present paper is the prediction of an additional 
dozen higher excited states obtained in the VMC calculations as shown in
Table~\ref{table:excite}, and in Fig.~\ref{fig:ex_spectra}.
We have calculated VMC and GFMC excitation energies for eight states, and they
agree with each other within error bars in all cases.
Therefore, we may expect that the VMC excitation energies for the other 
states shown are close to the correct results for this Hamiltonian.
Most of these higher states are obtained by the diagonalizations
within correlated $p$-shell states discussed in Sec.III.

In Fig.~\ref{fig:energy_spectra} we see that the difference between
the calculated and experimental energies increases
as $A$ and $|N-Z|$ increase.
However, as seen in Fig.~\ref{fig:ex_spectra},
the excitation spectra are in good overall agreement with experiment.
The states generally occur in the correct order, and with reasonable energies.
The agreement with the $^6$Li and low-lying $^7$Li spectra is very good.
In particular, the first excited $(1^+;0)$ state in $^6$Li, and the first
excited $(\case{3}{2}^-;\case{1}{2})$ state in $^7$Li, i.e., the first excited
states with quantum numbers identical to the ground states, are very close
to the observed excitations.
In the case of $^6$He and $^7$He, we predict a number of states that have
not been observed experimentally, but which could be searched for.  
A first observation of the second $(2^+;1)$ state in $^6$He was recently 
reported~\cite{MSU96}; the experimenters tried to fit their data with single
states of different $(J^{\pi};T)$ but did not get a very good fit for any one
value.
Our results suggest there are several states close together in this region,
which could improve the chances of fitting the data satisfactorily. 
The states in $^7$He might also be amenable to experimental measurement with
the new radioactive beam facilities that are now coming on line.

\subsection{Isobaric analog states}

Energy differences of isobaric analog states are sensitive probes of the
charge-independence-breaking parts of the Hamiltonian.
To study these it is useful to express the energies in an isobaric multiplet,
characterized by $A$ and $T$, in terms of the isospin multipole operators
of order $n$:
\begin{equation}
   E_{A,T}(T_z) = \sum_{n\leq 2T} a^{(n)}_{A,T} Q_n(T,T_z) \ .
\end{equation}
The $Q_n(T,T_z)$ are orthogonal functions for projecting out isovector,
isotensor, and higher-order terms~\cite{P60}; the first terms are
$Q_0=1$, $Q_1=T_z$, and $Q_2=\case{1}{2}(3T_z^2-T^2)$.
The coefficients $a^{(n)}$ are then obtained from
\begin{equation}
   a^{(n)}_{A,T} = \sum_{T_z} Q_n(T,T_z) E_{A,T}(T_z) 
                 / \sum_{T_z} Q^2_n(T,T_z) \ .
\end{equation}

In first-order perturbation theory, the electromagnetic interaction
contributes to the $a^{(n)}$ for $n=1$ and 2, the nuclear CSB potential
and kinetic energy
contribute to $n=1$, and the nuclear CD potential contributes to $n=2$.
The $a^{(n)}$ for higher $n$ are zero in first order with our Hamiltonian, and 
there is little experimental evidence for $n\geq3$ terms in nuclei~\cite{BK79}.
We have made VMC calculations of the $a^{(1,2)}$ in first order by using
a CI wave function of good isospin, $T$, and simply varying $T_z$
to compute the $E_{A,T}(T_z)$.
Table~\ref{table:analog} contains results for the $T=\case{1}{2}$ isovector 
($n=1$) coefficients in $A=3$ and $A=7$, the $T=1$ isovector and isotensor 
($n=2$) coefficients in $A=6$, and the $T=\case{3}{2}$ isovector and isotensor
coefficients in $A=7$.
The energy differences are broken down into $v^{\gamma}$, $v^{CD}$, $v^{CSB}$, 
and $K^{CSB}$ contributions, with the $pp$ Coulomb, [$v_{C1}(pp)$], other 
Coulomb ($v_{CR}$), and magnetic moment ($v_{MM}$) components of $v^{\gamma}$
also given.

The CIB parts of the Hamiltonian induce CD changes in the nuclear 
wave function, leading to higher-order perturbative corrections to the 
splittings of the isospin multiplets.
We have estimated some of these changes in VMC by repeating the calculations
using wave functions that have a varying Coulomb term, $V^C_{\alpha N}$ of 
Eq.(\ref{eq:vcan}), added to the single-particle potential well that is used 
to generate the $\phi^{LS}_p(R_{\alpha k})$ components of $\Psi_V$.
This results in a slightly more diffuse wave function as $Z$ increases, and 
slightly smaller energy coefficients than those obtained with CI wave functions.
In the GFMC calculations, the isoscalar Coulomb term, $v^{\prime}_C$, 
provides an additional source of CIB through the propagating Hamiltonian of
Eq.(\ref{eq:hgfmc}), which depends on $T_z$.
However, within the limited propagation time of $\tau = 0.06$ MeV$^{-1}$,
the main effect of using GFMC wave functions seems to be the slightly sharper
two-body densities (discussed below) around 1 fm, and consequent changes
in the CIB potential expectation values.
All our GFMC calculations have been made with CD wave functions, but a complete
set of isobaric analog states was calculated only for $A$=6, results for
which are shown on the penultimate line of Table~\ref{table:analog}.
There may also be higher-order contributions to the isomultiplet splittings
from changes to the CI expectation values, but we have no reliable way of 
extracting these from under the sizeable Monte Carlo errors.

The results tabulated in Table~\ref{table:analog} indicate that the present
Hamiltonian underestimates the observed isovector coefficients and gives mixed
results for the isotensors.
It should be remembered, however, that while $v^{CD}$ is well determined in 
the $\ell=0$ partial wave by the $N\!N$ scattering data, it is much less 
well known in $\ell=1$ and higher partial waves, while the only experimental
input for $v^{CSB}$ is the $nn$-$pp$ scattering length difference, which has 
a 20\% experimental uncertainty.
The $A=3$ case would be corrected by a $\sim 10\%$ increase in $v^{CSB}$, but this
would not explain much of the discrepancy in the larger nuclei.
The significant underbinding of the $A=6,7$ nuclei with the present Hamiltonian 
may mean that our wave functions for these nuclei are more diffuse than they 
should be.
By far the worst discrepancy is for the $A,T=7,\case{3}{2}$ case, where the
underbinding of the ground states is also the largest.
If the Hamiltonian were altered, e.g., by increasing the net attraction from
the three-nucleon interaction, to obtain the correct binding, the contribution
of the CIB forces to the isovector coefficients should be increased, both in
the dominant $pp$ Coulomb term, and the short-range $v^{CSB}$.
It is more difficult to predict the effect of such changes on the isotensor 
energy coefficients.

\section{One- and Two-Nucleon Distributions}

   The one- and two-nucleon distributions of light $p$-shell nuclei
are interesting in a variety of experimental settings.  
For example, the $^6$He nucleus has been a popular candidate for study as 
a 'halo' nucleus whose last two neutrons are weakly bound.  
In addition, the polarization densities of $^6$Li and $^7$Li are important 
because of possible applications in polarized targets.  
In order to extract information on the spin-dependent nucleon properties 
from experiments on such targets one must, at a minimum, understand
the nucleon polarization in the polarized nucleus.  
In this section we provide our results for a variety of nucleon distributions, 
including spin-polarized and averaged single-nucleon densities, spin-dependent
and spin-independent two-body densities, the proton-proton distributions,
and the rms radii, magnetic moments, and quadrupole moments.

As was discussed in Sec.~VI, we do not propagate to large enough
imaginary time to allow the GFMC to significantly modify the
rms radius of the $p$-shell nuclei.  
Thus they are determined almost entirely by the input trial wave function,
which is constrained to be near the experimental value wherever known.
However GFMC does make significant changes to densities at small $r$.
A number of the one-body densities are increased near the origin,
as are the peaks of many of the two-body densities.

The proton rms radii and static electromagnetic properties are given in 
Table~\ref{table:radii}.
These are calculated from $\Psi_V$ using impulse approximation.
In general, we know that there are significant corrections to the 
electromagnetic moments from two-body charge and current 
contributions~\cite{SPR89,SPR90}.
For the magnetic moments, these corrections are only $\sim$ 1-2\% in 
isoscalar nuclei like $^2$H, but are $\sim$ 15-20\% in the isovector
$T=\case{1}{2}$ nuclei $^3$H and $^3$He.
Therefore it is not surprising that we see very little discrepancy for the 
magnetic moment in $^6$Li, but a sizeable error for $^7$Li; presumably, a 
calculation including meson-exchange contributions would come much closer 
to the experimental values in the latter case.
The quadrupole moment is a more difficult problem, particularly in $^6$Li,
where there is a delicate cancellation between the contributions from the
deuteron quadrupole moment and the $D$-wave part of the $\alpha$-$d$
relative wave function.
Many cluster models for $^6$Li fail to obtain the observed negative sign;
we have trouble getting an accurate measure of the magnitude, for 
reasons discussed above.
In the case of $^7$Li, where there is no such delicate cancellation, 
the value is only $\sim$ 15-20\% too low in magnitude. 
Again, some of this discrepancy might be made up by meson-exchange corrections.

In Fig.~\ref{fig:np_dens_he}, we present the neutron and proton densities for 
the helium isotopes, calculated with GFMC.  
Previously we have found that the $^4$He charge form factor is in good 
agreement with experimental data in realistic calculations~\cite{SPR90}, 
and hence this distribution should be quite accurate.  
As more neutrons are added, the tails of the distributions broaden considerably 
because of the relatively weak binding of the $p$-shell neutrons.  
In addition, the central neutron and proton densities decrease rather 
dramatically.  
This effect does not necessarily require any changes to the alpha-particle 
core, but can be understood at least partially from the fact that the 
alpha particle no longer sits at the center of mass of the entire system.  
The motion relative to the center of mass spreads out the mass distribution 
relative to that of $^4$He.

We also find that the small depression obtained in the central
density of $^4$He gradually disappears as more nucleons are added.
While the depression is clear in the alpha particle, it is
nearly within our statistical errors in $^6$He and seems to have
disappeared completely in $^7$He. This can again be understood
by taking into account the fluctuations in the center of mass
of the core nucleons about the center of mass of the entire system.
Fig.~\ref{fig:np_dens_li} shows the GFMC neutron and proton 
densities for the lithium
isotopes. These densities are very smooth functions of the distance
from the center of mass. 

The polarization densities for $^6$Li and $^7$Li, computed with VMC
are presented in Figs.~\ref{fig:spin_dens_li6}  
and \ref{fig:spin_dens_li7}, respectively.  
The spin-up proton density distribution is defined by
\begin{equation}
\rho_{p\uparrow}(r) \ = \ \frac{1}{4 \pi r^2}
\langle \Psi ( J, M_J = J ) | \sum_i \ \frac{1 + \sigma_z (i)}{2}
\ \frac { 1 + \tau_z (i)}{2} \delta ( r - | {\bf r}_i - {\bf R}_{cm} | ) 
\ | \Psi ( J, M_J = J ) \rangle \,
\end{equation}
with similar definitions for spin-down protons, spin-up neutrons, etc.
The integral of these distributions is the total number
of spin-up (down) protons (neutrons) in a fully polarized state.
The integrated quantities can be important
in high-energy experiments designed to probe the spin-dependence
of the neutron or proton structure functions, while their radial dependence
may be partially accessible in experiments at lower energies.
Experimentalists are considering using dense, solid polarized $^6$LiD targets,
as an alternative to the deuterated ammonia targets currently being used to 
probe neutron properties.

Polarization densities have been studied previously
in cluster models, with a fixed (unpolarized) alpha core
plus interacting valence nucleons.  Our calculations include
the possibility of the spins in the the core alpha particle being
polarized by the valence nucleon's spin and orbital angular momentum, and hence
it is interesting to examine the results for both the distributions
and the integrated quantities.

In the spin projection $M=1$ state of the deuteron, the polarization of the 
neutron differs from unity because the tensor interaction induces a D-state
in the wave function. Integrating $\rho_{n\uparrow}(r)$ over $r$ yields
a probability for up-spin neutrons of
\begin{equation}
P(n\!\!\uparrow) = P_S + \case{1}{4} P_D = 1 - \case{3}{4} P_D ~,
\end{equation}
where $P_S$ and $P_D$ are the S- and D-wave probabilities of the
deuteron.
In the simplest two-body (alpha plus deuteron) model of 
$^6$Li, the polarization of the neutrons is determined
by the D-state probability in the deuteron and
the D-state probability in the $\alpha-d$ wave function.
In three-body models, recent calculations~\cite{SKCA93} have found
that the valence neutron had $P(n\!\!\uparrow) = 0.93(1)$.
Since in such a model the two core neutrons are unpolarized,
this corresponds to a total projection $P (n\!\!\uparrow)$ of 1.93,
or a polarization of 29\%.

As expected, the up spins dominate the down spins in the $M=1$ state of 
$^6$Li at large distances from the center of mass.  
At very large separations, the ratio will be determined solely by the 
asymptotic D/S state normalization of the alpha-deuteron wave function and the 
$D$-state probability in the deuteron.
At small r, we find that the spin-down density exceeds the spin-up
density, presumably because the spins of the outer nucleons
prefer to try to pair with the core nucleons to a spin-zero state.
Even though we find this effect to be significant, the integrated
spin densities agree reasonably well with the cluster model calculations.  
The integrated neutron densities in $^6$Li are found to be 1.93 for spin up and 
1.07 for spin down, respectively, yielding the same net polarization of 29\%.

The $P(p\!\!\uparrow)$ in $^7$Li is found to be 1.94 instead of 2 
as predicted by the independent-particle shell model.
The neutrons carry about half this remaining spin, as the spin-up
and spin-down neutron integrated densities are 1.98 and 2.02.
 
We have also computed a variety of two-nucleon distribution functions.
These are defined by
\begin{equation}
\rho_{2,p} (r) = \frac{1}{4 \pi r^2}
\langle \Psi | \sum_{i<j}  O^p_{ij} \delta ( r - |{\bf r}_i - {\bf r}_j|)|
\Psi \rangle ~,
\end{equation}
where the operators $O^k_{ij}$ are given in Eq.(\ref{eq:osix}).
While typically these two-body correlations cannot be directly measured,
they provide the expectation values of two-body operators and
can be important ingredients in interpreting the results of experiments.
Transition matrix elements of this type are needed for
extracting the effective weak $\pi N\!N$ coupling
constant in parity-violating experiments.
Fig.~\ref{fig:li6_ttdenr} 
shows VMC and GFMC calculations of the $S_{ij} \tau_i \cdot \tau_j$ 
$N\!N$ pair distribution function.
This correlation is strongly dominated by pion exchange,
and is responsible for the toroidal shapes
which characterize the coupling of spin to space in the 
nucleus\cite{FPPWSA96}.  We see that the structure is somewhat
enhanced by the GFMC.

Finally, we present the proton-proton distributions 
(scaled to have normalization integrals of $Z\!-\!1$) for
$^4$He, $^6$He, $^7$He, $^6$Li, and $^7$Li in Fig.~\ref{fig:ppcorr}.
These distributions are directly related to the Coulomb sum 
measured in inclusive longitudinal electron scattering;
such measurements in $^3$He have been used\cite{Beck} to
put constraints on the $\rho_{pp} (r_{ij})$, and realistic calculations
agree with the experimental results\cite{SWC93}.
The behavior of $\rho_{pp} (r)$ at short distances is largely determined
by the repulsive core of the $N\!N$ potential and is nearly independent
of the nucleus, but at larger distances it is determined by the size 
of the nucleus.

We show results for $\rho_{pp}$ in $^6$He and $^7$He in order to directly
compare the alpha particle proton-proton distribution
to that in the alpha-particle cores
of $^6$He and $^7$He.  Unlike the one-body densities, these distributions are
not sensitive to center of mass effects.
We find that the proton-proton distribution spreads out slightly with
neutron number in the helium isotopes, with an increase of the pair rms radius of
approximately 4\% in going from $^4$He to  $^6$He, and 7\% to $^7$He.
While this could be interpreted as a swelling of the alpha core,
it might also be due to the charge-exchange ($\tau_i \cdot \tau_j$) correlations
which can transfer the charge from the core to the valence nucleons.
Since these correlations are rather long-ranged, they can have
a significant effect on the proton-proton distribution.

Finally, we mention that calculations of the electromagnetic
ground-state and transition form factors are underway.
Complete results for these quantities, including exchange
current effects,  will appear in a separate paper.

\section{CONCLUSIONS}

Quantum Monte Carlo methods are now a powerful tool for the study of
light $p$-shell nuclei.
At present, we can write plausible variational wave functions with the proper
quantum numbers for the given state of interest, but they do not give 
sufficient binding to provide stability against breakup into subclusters.
However, the GFMC method rapidly damps out the small amount of highly-excited
states contained in the VMC wave functions, producing ground states that
are stable in the case of $^6$Li and $^7$Li.
The current major limitation is the small value of $\tau$ that can be reached
in most calculations.
This makes it important that the starting VMC wave functions have a proper mix 
of the appropriate spatial symmetries, and negligible contamination from
low-energy excited states.

The energies obtained for the ground and low-lying excited states are close to,
but somewhat above, the experimental numbers.
We believe the discrepancy is probably the fault of the Hamiltonian,
most likely the phenomenological short-range part of the three-nucleon 
interaction, rather than a shortcoming of the calculation.
We note that the difference between experimental and theoretical energies is
much less that $\langle V_{ijk} \rangle$, and might be rectified by 
an improved three-nucleon potential.
Despite the discrepancies in the ground state energies, the excitation spectra
are reproduced very well.
We believe this is the first demonstration that the shell structure of light
nuclei can be obtained directly from bare two-nucleon interactions that fit 
$N\!N$ scattering data.

The QMC methods developed here can be extended to eight-body nuclei with the
present generation of computers.
We already have calculations in progress for the ground and low-lying excited
states of $A=8$ nuclei.
The next major task will be to refine our model for the three-nucleon
interaction, perhaps including those relativistic corrections which first
appear at the three-nucleon level~\cite{FPF95,FPCR95}, with the intention of 
fitting the energies of $A$=3-8 nuclear states with 1\% accuracy.
Now that accurate QMC calculations of these states are possible, there are a 
host of interesting problems that become accessible, including the response of 
$^6$Li and $^7$Li to electron scattering, and many low-energy electroweak
capture reactions of astrophysical interest, such as $^4$He$(d,\gamma)^6$Li
and $^7$Be$(p,\gamma)^8$B.
There also remains the problem of adapting the GFMC methods here to the study 
of larger systems, perhaps through methods similar to the cluster-expansion
used in VMC calculations of $^{16}$O~\cite{PWP92}.

\acknowledgements
The authors thank Dr. Dieter Kurath for many useful suggestions.
The many-body calculations were performed on the IBM SP of the Mathematics
and Computer Science Division, Argonne National Laboratory, and on the IBM SP2
of the Cornell Theory Center.
The work of BSP and VRP is supported by the U.S. National Science Foundation
via grant PHY89-21025, that of JC by the U.S. Department of Energy, and that
of SCP and RBW by the U.S.  Department of Energy, Nuclear Physics Division, 
under contract No. W-31-109-ENG-38.

\narrowtext
\begin{table}
\caption{Values of shape parameters used in generation of trial functions.
All units in fm.
Notation is same as Ref.~\protect\cite{W91}.}
\begin{tabular}{ldldld}
  $a_0$    &  0.35  & $c_0$    &  1.1   & $R_0$    &  0.75  \\
  $a_1$    &  0.4   & $c_1$    &  3.0   & $R_1$    &  2.8   \\
  $a_t$    &  0.4   & $d$      &  2.0   & $R_t$    &  3.7   \\
  $a_b$    &  0.24  &          &        & $R_b$    &  0.4   
\label{table:params1}
\end{tabular}
\end{table}

\begin{table}
\caption{Values of asymptotic parameters used in generation of trial functions.
Notation is same as Ref.~\protect\cite{W91}.}
\begin{tabular}{ldd}
              & $^3$H    & $A \geq 4$      \\
\tableline
  $E_{0,0}$ (MeV)   &   3.2    &  17.0 \\
  $E_{0,1}$ (MeV)   &   6.0    &  16.0 \\
  $E_{1,0}$ (MeV)   &  13.0    &  23.5 \\
  $E_{1,1}$ (MeV)   &   6.4    &  16.5 \\
  $h_{0,0}$         &   0.85   &   1.04  \\
  $h_{0,1}$         &   1.70   &   1.71  \\
  $h_{1,0}$         &   1.74   &   1.54  \\
  $h_{1,1}$         &   1.72   &   1.68  \\
  $\eta_0$          &   0.026  &   0.035 \\
  $\eta_1$          & --0.007  & --0.015 \\
  $\zeta_1$         &   0.0003 &   0.0003 \\
  $\alpha_t$        &   0.92   &   0.86
\end{tabular}
\label{table:params2}
\end{table}

\begin{table}
\caption{Values of three-nucleon correlation parameters.
Notation is same as Ref.~\protect\cite{APW95}.}
\begin{tabular}{ld}
  $\epsilon_a$                 &  --0.00025    \\
  $\epsilon_c$                 &  --0.0004     \\
  $\epsilon_u$                 &  --0.0005     \\
  $y$                          &    0.72       \\
  $q^{c}_{1}$ (fm$^{-6}$)      &    0.20      \\
  $q^{c}_{2}$ (fm$^{-1}$)      &    0.85      \\
  $q^{p}_{1}$                  &    0.16       \\
  $q^{p}_{2}$ (fm$^{-1}$)      &    0.05      \\
  $q^{\ell s}_{1}$ (fm$^{-2}$) &  --0.12      \\
  $q^{\ell s}_{2}$ (fm$^{-2}$) &    0.12      \\
  $q^{\ell s}_{3}$ (fm$^{-2}$) &    0.85      \\
  $q^{\tau}_{1}$ (fm$^{-1}$)   &  --0.012     \\
  $q^{\tau}_{2}$ (fm$^{-2}$)   &    0.015     \\
  $q^{\tau}_{3}$ (fm)          &    1.2       \\
  $q^{\tau}_{4}$               &    0.35    
\end{tabular}
\label{table:params3}
\end{table}

\begin{table}
\caption{Energy obtained with different trial functions for $^6$Li in MeV, and
relative cost to compute.}
\begin{tabular}{lddd}
  wave function             &   $E_V$     & $\delta E$ & cost \\
  $|\Psi_P\rangle$          & --25.47(30) &            & 1.00 \\
  $|\Psi_T\rangle$          & --27.00(13) &   --1.53   & 1.19 \\
  $[1 + \sum_{i<j<k}U^{TNI}_{ijk}] |\Psi_P\rangle$ 
                            & --27.11(12) &   --1.64   & 1.59 \\
  $[1 + \sum_{i<j}U^{LS}_{ij} + \sum_{i<j<k}U^{TNI}_{ijk}] |\Psi_P\rangle$ 
                            & --27.89(12) &   --2.42   & 2.23 \\
  $|\Psi_V\rangle$          & --28.14(11) &   --2.67   & 2.66
\end{tabular}
\label{table:psiv}
\end{table}

\begin{table}
\caption{Values of variational parameters in $p$-shell nuclei.}
\begin{tabular}{ldddddd}
&\multicolumn{3}{c}{$A=6$}&\multicolumn{3}{c}{$A=7$}             \\
& $^{1,3}$S[2] & $^{1,3}$D[2] & $^{1,3}$P[11] & $^2$P[3] & $^2$F[3] 
& [21] \& [111] \\
\tableline
  $a_{sp}$
&    0.0     &    0.0     &    0.0     &    1.0     &    1.0     &    1.0     \\
  $b_{sp}$
&    1.0     &    1.0     &    1.0     &    0.0     &    0.0     &    0.0     \\
  $c_{sp}$
&    0.9     &    0.9     &    0.9     &    0.85    &    0.85    &    0.85    \\
  $d_{sp}$ (fm)
&    3.2     &    3.2     &    3.2     &    3.2     &    3.2     &    3.2     \\
  $a_{pp}$
&    1.0     &    1.0     &    1.0     &    1.0     &    1.0     &    1.0     \\
  $b_{pp}$
&    0.0     &    0.0     &    0.0     &    0.0     &    0.0     &    0.0     \\
  $c_{pp}$
&    0.1     &    0.1     &    0.4     &    0.1     &    0.1     &    0.4     \\
  $d_{pp}$ (fm)
&    3.2     &    3.2     &    3.2     &    3.2     &    3.2     &    3.2     \\
  $R_{f}$ (fm)
&    4.0     &    4.0     &    4.0     &    4.0     &    4.0     &    4.0     \\
  $a_{f}$ (fm)
&    1.0     &    1.0     &    1.0     &    1.0     &    1.0     &    1.0     \\
  $V_p$ (MeV) 
& --20.0     & --18.0     & --18.0     & --20.0     & --18.0     & --18.0     \\
  $R_p$ (fm)  
&    4.0     &    4.0     &    4.0     &    4.0     &    4.0     &    4.0     \\
  $a_p$ (fm) 
&    1.5     &    1.5     &    1.5     &    1.5     &    1.5     &    1.5     \\
\end{tabular}
\label{table:params4}
\end{table}

\begin{table}
\caption{$\beta_{LSn}$ components in $A=6$ states, listed in order of 
increasing excitation for $T=0$ and $T=1$.}
\begin{tabular}{cdddddd}
$(J^{\pi};T)$ & 
         $^1$S[2] & $^3$S[2] & $^1$D[2] & $^3$D[2] & $^1$P[11] & $^3$P[11] \\
$(1^+,0)$ &  ---  &    0.987 &   ---    & 0.117    &   0.111   &   ---   \\
$(3^+;0)$ &  ---  &    ---   &   ---    & 1.       &   ---     &   ---   \\
$(2^+;0)$ &  ---  &    ---   &   ---    & 1.       &   ---     &   ---   \\
$(1^+,0)$ &  ---  &  --0.074 &   ---    & 0.949    & --0.306   &   ---   \\
$(1^+,0)$ &  ---  &  --0.153 &   ---    & 0.300    &   0.942   &   ---   \\
\tableline
$(0^+;1)$ & 0.967 &    ---   &   ---    & ---      &   ---     & --0.253 \\
$(2^+;1)$ &  ---  &    ---   &   0.880  & ---      &   ---     &   0.476 \\
$(2^+;1)$ &  ---  &    ---   & --0.476  & ---      &   ---     &   0.878 \\
$(1^+;1)$ &  ---  &    ---   &   ---    & ---      &   ---     &   1.    \\
$(0^+;1)$ & 0.262 &    ---   &   ---    & ---      &   ---     &   0.965 \\
\end{tabular}
\label{table:beta6}
\end{table}

\begin{table}
\caption{$\beta_{LSn}$ components in $A=7$ states, listed in order of 
increasing excitation for $T=\case{1}{2}$ and $T=\case{3}{2}$.}
\begin{tabular}{cdddddddd}
$(J^{\pi};T)$ & 
 $^2$P[3] & $^2$F[3] & $^2$P[21] & $^4$P[21] & 
                              $^2$D[21] & $^4$D[21] & $^2$S[111] & $^4$S[111] \\
$(\case{3}{2}^-,\case{1}{2})$ 
&   0.998 &   ---   &   0.001 & 0.050 & --0.041 &   0.012 &   ---   &   ---   \\
$(\case{1}{2}^-,\case{1}{2})$ 
&   0.994 &   ---   & --0.087 & 0.001 &   ---   & --0.068 & --0.010 &   ---   \\
$(\case{7}{2}^-,\case{1}{2})$ 
&   ---   &   0.998 &   ---   & ---   &   ---   &   0.059 &   ---   &   ---   \\
$(\case{5}{2}^-,\case{1}{2})$ 
&   ---   &   0.995 &   ---   & 0.073 & --0.060 &   0.036 &   ---   &   ---   \\
$(\case{5}{2}^-,\case{1}{2})$ 
&   ---   & --0.059 &   ---   & 0.969 &   0.168 & --0.171 &   ---   &   ---   \\
$(\case{7}{2}^-,\case{1}{2})$ 
&   ---   & --0.052 &   ---   & ---   &   ---   &   0.999 &   ---   &   ---   \\
$(\case{3}{2}^-,\case{1}{2})$ 
& --0.041 &   ---   & --0.022 & 0.998 &   0.039 & --0.015 &   ---   &   ---   \\
$(\case{1}{2}^-,\case{1}{2})$ 
&   0.035 &   ---   &   0.412 & 0.909 &   ---   & --0.014 & --0.057 &   ---   \\
\tableline
$(\case{3}{2}^-,\case{3}{2})$ 
&   ---   &   ---   &   0.864 & ---   &   0.480 &   ---   &   ---   & --0.153 \\
$(\case{1}{2}^-,\case{3}{2})$ 
&   ---   &   ---   &   1.    & ---   &   ---   &   ---   &   ---   &   ---   \\
$(\case{5}{2}^-,\case{3}{2})$ 
&   ---   &   ---   &   ---   & ---   &   1.    &   ---   &   ---   &   ---   \\
$(\case{3}{2}^-,\case{3}{2})$ 
&   ---   &   ---   & --0.448 & ---   &   0.841 &   ---   &   ---   &   0.303 \\
\end{tabular}
\label{table:beta7}
\end{table}

\begin{table}
\caption{GFMC program performance on IBM SP1 and SP2 Wide Nodes. N(A,T) is the
number of spin-isospin states in the wave function and P is the number
of pairs.  The columns give the times for one propagation step and one
energy evaluation and the total time needed for a 50,000 configuration
calculation; see the text for a more complete description.}
\begin{tabular}{lrrddddrr}
 &N(A,T)&P&\multicolumn{2}{c}{Propagation}&\multicolumn{2}{c}{Energy Calculation}&
    \multicolumn{2}{c}{Total} \\
 & & &\multicolumn{2}{c}{msec.}&\multicolumn{2}{c}{sec.}&
    \multicolumn{2}{c}{node hours} \\
 & & & SP1~ & SP2~ & SP1 & SP2 & SP1 & SP2 \\
\tableline
$^{4}He$ &        32&  6& 6.6  & 3.4 & 0.025  & 0.01  & 13    & 7 \\
$A=6;~ T=0$ &     320& 15& 63.  & 26.&  0.84   &  0.31 & 220   & 80 \\
$A=6;~ T=1$ &     576& 15& 100. & 40.&  1.66   &  0.56 & 290   & 125 \\
$A=7;~ T=\case{1}{2},\case{3}{2}$ &
                1792& 21& 460. & 170.& 10.6  & 3.4  & 2,230 & 725 \\
\end{tabular}
\label{table:perform}
\end{table}

\begin{table}
\caption{Contributions to the GFMC $\langle O(\tau) \rangle$ of Eq.~\protect\ref{eq:pc_gfmc} for $^6$Li.  All quantities are in MeV.}
\begin{tabular}{ldddd}
 &                $\langle O \rangle_T$ & $\langle O \rangle_{Mixed}$ &
 $\langle O \rangle_{Mixed}-\langle O \rangle_T$ &$\langle O \rangle$ \\
\tableline
$K^{CI}$           &    143.8(4) &   147.3(5) &     3.5(7) &   150.8(10) \\
$v^{\prime}_{8}$   &  --168.7(4) & --175.7(6) &   --7.0(8) & --182.6(11) \\
$v^{\prime}_{C}$   &      1.5(0) &     1.5(0) &     0.0(0) &     1.5(0) \\
$V^{\prime}_{ijk}$ &    --3.5(1) &   --4.4(1) &   --0.9(1) &   --5.4(1) \\
Sum                &   --26.9(1) &  --31.3(1) &   --4.4(1) &  --35.7(1) \\
$H^{\prime}$       &   --26.9(1) &  --31.3(1) &     --     &  --31.3(1) \\
\end{tabular}
\label{table:pc_cor}
\end{table}

\begin{table}
\caption{Experimental and quantum Monte Carlo energies of $A=2 - 7$ nuclei
in MeV}
\begin{tabular}{ldddd}
$^AZ(J^{\pi};T)$ & VMC ($\Psi_T$) & VMC ($\Psi_V$) &   GFMC    &   Expt   \\
\tableline
$^2$H$(1^+;0)$      &  --2.2248(5)&             &              &  --2.2246 \\
$^3$H$(\case{1}{2}^+;\case{1}{2})$  
                    &  --8.15(1)  &  --8.32(1)  &   --8.47(1)  &  --8.48  \\

$^4$He$(0^+;0)$     & --26.93(2)  & --27.76(3)  &  --28.30(2)  & --28.30  \\

$^6$He$(0^+;1)$     & --23.77(6)  & --24.87(7)  &  --27.64(14) & --29.27  \\
$^6$He$(2^+;1)$     & --22.05(6)  & --23.01(7)  &  --25.84(11) & --27.47  \\

$^6$Li$(1^+;0)$     & --27.04(3)  & --28.09(7)  &  --31.25(11) & --31.99  \\
$^6$Li$(3^+;0)$     & --23.98(7)  & --25.16(7)  &  --28.53(32) & --29.80  \\
$^6$Li$(0^+;1)$     & --23.18(6)  & --24.25(7)  &  --27.31(15) & --28.43  \\
$^6$Li$(2^+;0)$     & --22.58(10) & --23.86(8)  &  --26.82(35) & --27.68  \\

$^6$Be$(0^+;1)$     & --21.73(6)  & --22.79(7)  &  --25.52(11) & --26.92  \\

$^7$He$(\case{3}{2}^-;\case{3}{2})$ 
                    & --19.02(8)  & --20.43(12) &  --25.16(16) & --28.82  \\

$^7$Li$(\case{3}{2}^-;\case{1}{2})$ 
                    & --31.59(8)  & --32.78(11) &  --37.44(28) & --39.24  \\
$^7$Li$(\case{1}{2}^-;\case{1}{2})$ 
                    & --31.13(8)  & --32.45(11) &  --36.68(30) & --38.76  \\
$^7$Li$(\case{7}{2}^-;\case{1}{2})$ 
                    & --25.77(6)  & --27.30(11) &  --31.72(30) & --34.61  \\
$^7$Li$(\case{5}{2}^-;\case{1}{2})$ 
                    & --24.91(7)  & --26.14(11) &  --30.88(35) & --32.56  \\
$^7$Li$(\case{3}{2}^-;\case{3}{2})$ 
                    & --18.27(7)  & --19.73(12) &  --24.79(18) & --28.00  \\
\end{tabular}
\label{table:energy}
\end{table}

\begin{table}
\caption{Kinetic and potential energy contributions to GFMC energies in MeV}
\begin{tabular}{ldddddd}
$^AZ(J^{\pi};T)$ & 
$K$ & $v_{ij}$ & $V_{ijk}$ & $v^{\gamma}_{ij}$ & $v^{\pi}_{ij}$ & $V^{2\pi}_{ijk}$ \\
\tableline
$^2$H$(1^+;0)$ & 
 19.81    &  --22.05    &   0.0    & 0.018   &  --21.28   &   0.0     \\
$^3$H$(\case{1}{2}^+;\case{1}{2})$ & 
 50.0(8)  &  --57.6(8)  & --1.20(7)& 0.04    &  --43.8(2) &  --2.2(1) \\
$^4$He$(0^+;0)$ & 
112.1(8)  & --136.4(8)  & --6.5(1) & 0.86(1) &  --99.4(2) & --11.8(1) \\
$^6$He$(0^+;1)$ & 
140.3(15) & --165.9(15) & --7.2(2) & 0.87(1) & --109.0(4) & --13.6(2) \\
$^6$He$(2^+;1)$ & 
131.9(14) & --155.7(13) & --7.0(1) & 0.86(1) & --106.2(5) & --13.1(2) \\
$^6$Li$(1^+;0)$ & 
150.8(10) & --180.9(10) & --7.2(1) & 1.71(1) & --128.9(5) & --13.7(3) \\
$^6$Li$(3^+;0)$ & 
146.7(29) & --174.4(31) & --7.1(2) & 1.71(2) & --119.9(5) & --13.9(4) \\
$^6$Li$(0^+;1)$ &
135.1(16) & --161.4(16) & --6.9(2) & 1.65(1) & --108.5(4) & --12.9(2) \\
$^6$Li$(2^+;0)$ & 
139.6(32) & --166.0(34) & --6.7(3) & 1.66(3) & --119.2(5) & --12.4(4) \\
$^6$Be$(0^+;1)$ &
134.8(16) & --160.5(16) & --6.8(2) & 2.97(2) & --108.0(4) & --12.8(2) \\
$^7$He$(\case{3}{2}^-;\case{3}{2})$ & 
146.0(17) & --171.2(17) & --7.4(2) & 0.86(1) & --109.9(6) & --14.1(2) \\
$^7$Li$(\case{3}{2}^-;\case{1}{2})$ & 
186.4(28) & --222.6(30) & --8.9(2) & 1.78(2) & --152.5(7) & --17.1(4) \\
$^7$Li$(\case{1}{2}^-;\case{1}{2})$ & 
183.0(32) & --219.1(35) & --8.2(3) & 1.76(2) & --151.5(7) & --16.1(4) \\
$^7$Li$(\case{7}{2}^-;\case{1}{2})$ & 
178.4(28) & --209.6(30) & --8.5(3) & 1.78(2) & --142.2(7) & --16.1(4) \\
$^7$Li$(\case{5}{2}^-;\case{1}{2})$ & 
169.1(31) & --200.2(33) & --7.1(3) & 1.73(2) & --143.2(7) & --14.2(4) \\
$^7$Li$(\case{3}{2}^-;\case{3}{2})$ & 
147.8(15) & --173.8(15) & --7.2(2) & 1.68(1) & --109.4(6) & --13.9(2)
\end{tabular}
\label{table:detail}
\end{table}

{\tighten
\begin{table}
\caption{Experimental, VMC and GFMC 
excitation energies (adjusted to their respective ground states) in MeV}
\begin{tabular}{lddddd}
$^AZ(J^{\pi};T)$                  & Experiment &  ~~~~VMC  & ~~~~GFMC \\
\tableline

\\                  
$^6$He$(2^+;1)$                     &  1.80 &  1.86(10) &  1.80(18) \\
$^6$He$(2^+;1)$                     &  5.6  &  3.61(10) & \\
$^6$He$(1^+;1)$                     &   ?   &  3.46(10) & \\
$^6$He$(0^+;1)$                     &   ?   &  5.24(11) & \\
\\                  
$^6$Li$(3^+;0)$                     &  2.19 &  2.93(10) &  2.72(36) \\
$^6$Li$(0^+;1)$                     &  3.56 &  3.84(10) &  3.94(23) \\
$^6$Li$(2^+;0)$                     &  4.31 &  4.23(11) &  4.43(39) \\
$^6$Li$(2^+;1)$                     &  5.37 &  5.64(10) & \\
$^6$Li$(1^+;0)$                     &  5.65 &  5.68(11) & \\
$^6$Li$(1^+;0)$                     &   ?   &  8.96(11) & \\
\\
$^7$He$(\case{1}{2}^-;\case{3}{2})$ &   ?   &  0.90(16) & \\
$^7$He$(\case{5}{2}^-;\case{3}{2})$ &   ?   &  1.69(16) & \\
$^7$He$(\case{3}{2}^-;\case{3}{2})$ &   ?   &  2.08(16) & \\
\\                                  
$^7$Li$(\case{1}{2}^-;\case{1}{2})$ &  0.48 &  0.33(16) &  0.76(41) \\
$^7$Li$(\case{7}{2}^-;\case{1}{2})$ &  4.63 &  5.48(16) &  5.72(41) \\
$^7$Li$(\case{5}{2}^-;\case{1}{2})$ &  6.68 &  6.64(16) &  6.56(45) \\
$^7$Li$(\case{5}{2}^-;\case{1}{2})$ &  7.46 &  9.90(16) & \\
$^7$Li$(\case{7}{2}^-;\case{1}{2})$ &  9.67 & 11.63(16) & \\
$^7$Li$(\case{3}{2}^-;\case{1}{2})$ &  9.90 & 10.14(16) & \\
$^7$Li$(\case{1}{2}^-;\case{1}{2})$ &   ?   & 10.79(16) & \\
$^7$Li$(\case{3}{2}^-;\case{3}{2})$ & 11.24 & 13.05(16) & 12.65(33) \\

\end{tabular}
\label{table:excite}
\end{table}
}

\begin{table}
\squeezetable
\caption{Breakdown of VMC isovector and isotensor energy coefficients
$a^{(n)}_{A,T}$ (in MeV) obtained with CI wave functions. 
Total coefficients are given for VMC with CI and CD wave functions, and for 
GFMC CD wave functions in $A$=6 nuclei.}
\begin{tabular}{ldddddd}
$A,T,n$ & 3,$\case{1}{2}$,1 & 6,1,1 & 6,1,2 & 7,$\case{1}{2}$,1 
        & 7,$\case{3}{2}$,1                 & 7,$\case{3}{2}$,2 \\
\tableline
$\langle v^{\gamma}\rangle$
& 0.680(1) & 1.048(2)  & 0.186(1)  & 1.501(3)  & 1.109(4)  & 0.119(1)  \\
$~~[\langle v_{C1}(pp)\rangle]$
&[0.651]   &[1.030]    &[0.167]    &[1.458]    &[1.099]    &[0.114]    \\
$~~[\langle v_{CR...}\rangle]$
&[0.011]   &[0.014]    &[0.001]    &[0.021]    &[0.012]    &[0.001]    \\
$~~[\langle v_{MM}\rangle]$
&[0.018]   &[0.004]    &[0.018]    &[0.023]    &[--0.002]  &[0.004]    \\
$\langle K^{CSB}\rangle$
& 0.014    & 0.014     & 0.        & 0.025     & 0.011     & 0.        \\
$\langle v^{CSB}\rangle$
& 0.066    & 0.035(1)  & 0.        & 0.080(1)  & 0.021(2)  & 0.        \\
$\langle v^{CD}\rangle$
& 0.       & 0.        & 0.101(12) & 0.        & 0.        & 0.020(4)  \\
$a^{(n)}_{A,T}$ (VMC: CI)
& 0.760(1) & 1.097(3)  & 0.287(12) & 1.605(4)  & 1.141(5)  & 0.139(4)  \\
$a^{(n)}_{A,T}$ (VMC: CD)
& 0.760(1) & 1.082(3)  & 0.277(12) & 1.597(4)  & 1.125(5)  & 0.132(4)  \\
$a^{(n)}_{A,T}$ (GFMC: CD)
& 0.756(1) & 1.120(9)  & 0.256(11) & ---       & ---       & ---       \\
$a^{(n)}_{A,T}$ (Expt.)
& 0.764    & 1.173     & 0.223     & 1.644     & 1.373     & 0.175
\end{tabular}
\label{table:analog}
\end{table}

\begin{table}
\caption{VMC values for proton rms radii (in fm), for quadrupole moments 
(in fm$^2$), and magnetic moments (in $\mu_N$) all in impulse approximation.
Only Monte Carlo statistical errors are shown; the limitation of GFMC 
propagation to $\tau = 0.06$~MeV${-1}$ introduces uncertainties of at least
0.1~fm in the rms radii.}
\begin{tabular}{ldddddd}
&\multicolumn{2}{c}{$\langle r^2_p \rangle^{1/2}$}
&\multicolumn{2}{c}{$\mu$}&\multicolumn{2}{c}{$Q$} \\
& VMC & experiment & VMC & experiment & VMC & experiment \\
\tableline
$^2$H$(1^+;0)$                      
& 1.967   & 1.953 &   0.847    &   0.857    &   0.270    &   0.286    \\

$^3$H$(\case{1}{2}^+;\case{1}{2})$  
& 1.59(1) & 1.60  &   2.582(1) &   2.979    &            &            \\

$^3$He$(\case{1}{2}^+;\case{1}{2})$ 
& 1.74(1) & 1.77  & --1.770(1) & --2.128    &            &            \\

$^4$He$(0^+;0)$                     
& 1.47(1) & 1.47  &            &            &            &            \\
 
$^6$He$(0^+;1)$                     
& 1.95(1) &       &            &            &            &            \\

$^6$Li$(1^+;0)$                     
& 2.46(2) & 2.43  &   0.828(1) &   0.822    & --0.33(18) & --0.083    \\

$^6$Be$(0^+;1)$                     
& 2.96(4) &       &            &            &            &            \\

$^7$Li$(\case{3}{2}^-;\case{1}{2})$ 
& 2.26(1) & 2.27  &   2.924(2) &   3.256    & --3.31(29) & --4.06     \\

$^7$Be$(\case{3}{2}^-;\case{1}{2})$ 
& 2.42(1) &       & --1.110(2) &            & --5.64(45) &            \\
\end{tabular}
\label{table:radii}
\end{table}

\begin{figure}
\caption{The experimental spectrum for $A=6$ nuclei.}
\label{fig:expt6}
\end{figure}

\begin{figure}
\caption{The experimental spectrum for $A=7$ nuclei.}
\label{fig:expt7}
\end{figure}

\begin{figure}
\caption{Statistical errors (MeV) in GFMC calculations with 50,000 initial
configurations as a function of imaginary time.}
\label{fig:sigma_of_tau}
\end{figure}

\begin{figure}
\caption{$^4$He GFMC energy as a function of imaginary time.  The fits
are described in the text.}
\label{fig:he4_e_tau}
\end{figure}

\begin{figure}
\caption{$^6$Li GFMC energy as a function of imaginary time.  The fits
are described in the text.}
\label{fig:li6_e_tau}
\end{figure}

\begin{figure}
\caption{$^6$Li GFMC energy as a function of imaginary time for
various truncations of the noncentral parts of $\Psi_T$.}
\label{fig:li6_trunc}
\end{figure}

\begin{figure}
\caption{$^6$He GFMC energy as a function of imaginary time for
$\Psi_T$ with various one-body $\Phi$.}
\label{fig:he6_phis}
\end{figure}


\begin{figure}
\caption{$^6$Li rms radii as a function of imaginary time for GFMC
calculations with $\Psi_T$ of varying rms radii.}
\label{fig:6li_rmsr}
\end{figure}

\begin{figure}
\caption{$^6$Li GFMC energies as a function of the GFMC rms radii from
calculations with $\Psi_T$ of varying rms radii.}
\label{fig:6li_e_vs_rmsr}
\end{figure}

\begin{figure}
\caption{$^4$He GFMC energy as a function of imaginary time for
the Argonne $v_{14}$ potential without $V_{ijk}$.  Also shown are several
previous calculations identified in the text.}
\label{fig:4he_v14}
\end{figure}

\begin{figure}
\caption{Spectrum for $A$=2-7 nuclei from experiment, and in GFMC and VMC
calculations.}
\label{fig:energy_spectra}
\end{figure}

\begin{figure}
\caption{$E(\tau)-E(\tau=0)$ for the ground states of $A$=3-7 nuclei.}
\label{fig:gfmc-vmc}
\end{figure}

\begin{figure}
\caption{$E(\tau)-E(0)/|E_{av}-E(0)|$ for the ground states of $A$=3-7 nuclei.}
\label{fig:normalized_gfmc-vmc}
\end{figure}

\begin{figure}
\caption{Excitation spectrum for $A$=6,7 nuclei from experiment, and in GFMC 
and VMC calculations.}
\label{fig:ex_spectra}
\end{figure}

\begin{figure}
\caption{The neutron and proton densities in $^4$He, $^6$He, and
$^7$He.}
\label{fig:np_dens_he}
\end{figure}

\begin{figure}
\caption{The neutron and proton densities in $^6$Li and
$^7$Li.}
\label{fig:np_dens_li}
\end{figure}

\begin{figure}
\caption{The spin-up and spin-down neutron and proton densities
in $^6$Li.}
\label{fig:spin_dens_li6}
\end{figure}

\begin{figure}
\caption{The spin-up and spin-down neutron and proton densities
in $^7$Li.}
\label{fig:spin_dens_li7}
\end{figure}

\begin{figure}
\caption{The two-nucleon $S_{ij} \tau_i \cdot \tau_j$ density of $^6$Li computed
from $\Psi_T$ and by GFMC.}
\label{fig:li6_ttdenr}
\end{figure}

\begin{figure}
\caption{The proton-proton densities in $^4$He, $^6$He, $^7$He,
$^6$Li, and $^7$Li nuclei.}
\label{fig:ppcorr}
\end{figure}

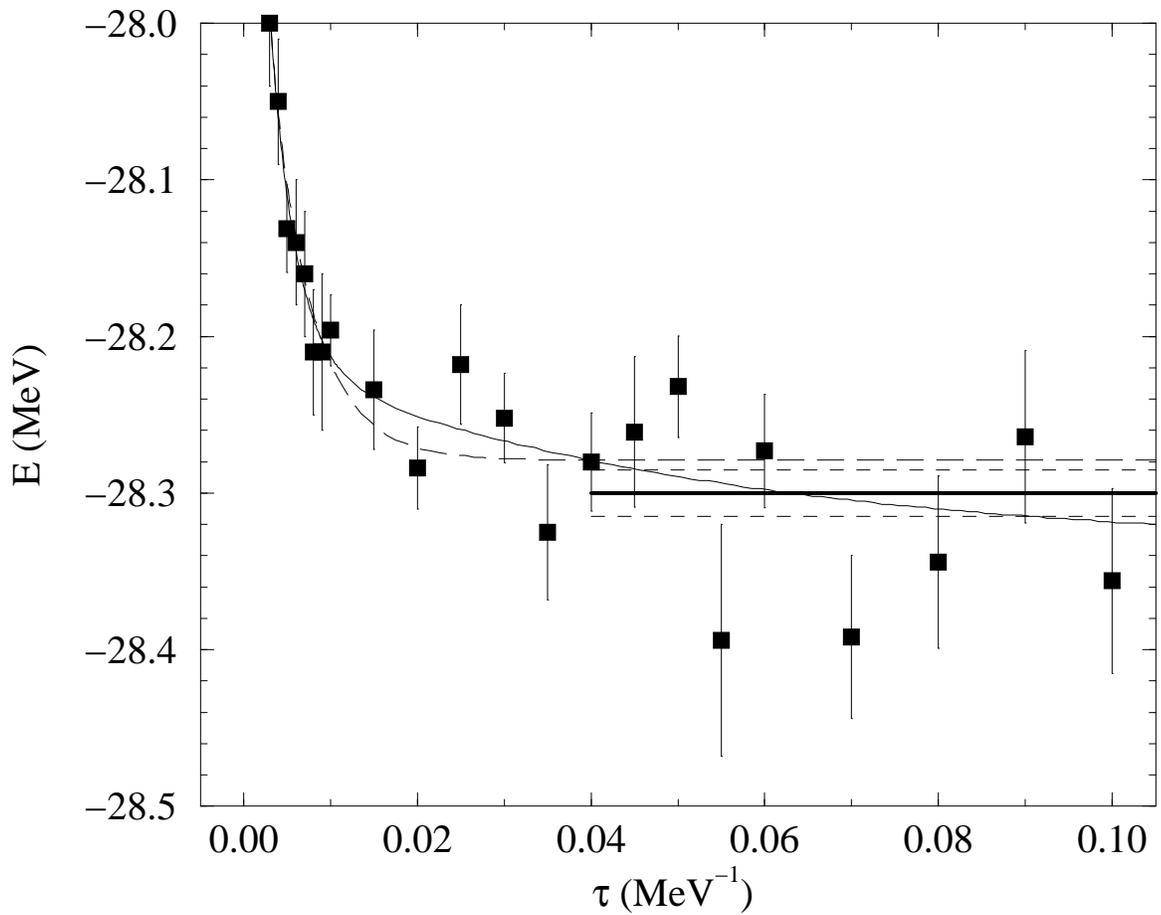
\end{document}